\documentclass[superscriptaddress,groupedaddress,nofootnoteinbib,11pt]{article}
\pdfoutput=1 
\usepackage{jheppub}

\usepackage{CJKutf8} 
\usepackage{mathtools,slashed,mathrsfs}
\usepackage[caption=false]{subfig}
\usepackage{dcolumn}
\usepackage{multirow}
\usepackage{tabularx}
\usepackage{booktabs}
\usepackage{bm}
\usepackage{comment}
\usepackage{setspace}
\usepackage[dvipsnames]{xcolor}
\usepackage[normalem]{ulem} 
\usepackage{enumerate}
\usepackage{siunitx}

\newcommand{\GeV}{{\, {\rm GeV}}}

\newcommand{\vUV}{v_\textsc{uv}}
\newcommand{\lUV}{\lambda_{h,\textsc{uv}}}
\newcommand{\TRH}{T_\textsc{rh}}
\newcommand{\MP}{M_P}
\newcommand{\LH}{\Lambda_{\mathcal H}}
\newcommand{\Lphi}{\Lambda_\phi}
\newcommand{\Lf}{\Lambda_f}
\newcommand{\fNL}{f_\mathrm{NL}}
\newcommand{\vk}{\vec k}
\newcommand{\vx}{\vec x}
\newcommand{\Pz}{\mathcal P_\zeta}
\newcommand{\oa}{\mathsf a}
\newcommand{\ob}{\mathsf b}
\newcommand{\oc}{\mathsf c}
\newcommand{\mt}{\widetilde{m}}
\newcommand{\mut}{\widetilde{\mu}}
\newcommand{\lt}{\widetilde{\lambda}}
\newcommand{\cth}{\cos\theta}
\newcommand{\sth}{\sin\theta}
\newcommand{\tf}{\widetilde{f}}
\definecolor{mypurple}{RGB}{164,64,214}

\newcommand\eea{\end{eqnarray}}
\newcommand\bea{\begin{eqnarray}}

\begin{document}

\title{Searches for other vacua II: A new Higgstory at the cosmological collider}

\date{\today}
\author[a]{Anson Hook,}
\author[b]{Junwu Huang,}
\author[b]{and Davide Racco}

\affiliation[a]{Maryland Center for Fundamental Physics, University of Maryland, College Park, MD 20742}
\affiliation[b]{Perimeter Institute for Theoretical Physics, 31 Caroline St.~N., Waterloo, Ontario N2L 2Y5, Canada}
\emailAdd{hook@umd.edu}
\emailAdd{jhuang@perimeterinstitute.ca}
\emailAdd{dracco@perimeterinstitute.ca}

\abstract{
The detection of an oscillating pattern in the bispectrum of density perturbations could suggest the existence of a high-energy second minimum in the Higgs potential.
If the Higgs field resided in this new minimum during inflation and was brought back to the electroweak vacuum by thermal corrections during reheating, the coupling of Standard Model particles to the inflaton would leave its imprint on the bispectrum.
We focus on the fermions, whose dispersion relation can be modified by the coupling to the inflaton, leading to an enhanced particle production during inflation even if their mass during inflation is larger than the Hubble scale.
This results in a large non-analytic contribution to non-Gaussianities, with an amplitude $\fNL$ as large as $100$ in the squeezed limit, potentially detectable in future 21-cm surveys. Measuring the contributions from two fermions would allow us to compute the ratio of their masses, and to ascribe the origin of the signal to a new Higgs minimum.
Such a discovery would be a tremendous step towards understanding the vacuum instability of the Higgs potential, and could have fascinating implications for anthropic considerations.
}

\maketitle

\section{Introduction and summary}

The discovery of the Higgs boson at the Large Hadron Collider (LHC) completes the Standard Model (SM) of particle physics.
Since then, much research has been done to understand the Higgs potential at both low and high energies.  
Extrapolating the predictions of the SM up to high energy scales, the quartic coupling of the Higgs becomes negative around $v_{\lambda=0}\sim 10^{11} \,{\rm GeV}$~\cite{Sher:1988mj,Arnold:1989cb,Altarelli:1994rb,Casas:1996aq,Hambye:1996wb,Isidori:2001bm,Ellis:2009tp,Bezrukov:2012sa,Bednyakov:2015sca,EliasMiro:2011aa,Degrassi:2012ry,Buttazzo:2013uya}. 
An epoch of primordial inflation, which would address many issues in cosmology 
\cite{Guth:1980zm,Starobinsky:1980te,Linde:1981mu,Mukhanov:1981xt,Albrecht:1982wi,Lyth:1998xn,Baumann:2009ds}, could have occurred at a high energy scale and can have a very interesting interplay with the Higgs instability~\cite{Espinosa:2007qp,Kobakhidze:2013tn,Kobakhidze:2014xda,Fairbairn:2014zia,Enqvist:2014bua,Hook:2014uia,Kamada:2014ufa,Shkerin:2015exa,Kearney:2015vba,East:2016anr,Herranen:2014cua}.
The common lore is that a future measurement of the scalar tensor ratio $r$~\cite{Abazajian:2016yjj} confirming high scale inflation would suggest that there is new physics below the scale $v_{\lambda=0}$ to stabilize the Higgs potential. 
In this paper, we take the opposite approach where we assume that during inflation the Higgs is living at a new minimum $\vUV$ at a scale well above $v_{\lambda=0}$.  
After inflation ends, the Higgs boson returns to the standard electroweak minimum due to thermal effects.  
In this article, we will explore the observational signatures associated with the Higgs living in its true minimum during inflation.

In the SM, the Higgs potential is unbounded from below.  In order to stabilize the potential, we assume that the potential is stabilized by higher dimensional operators. 
In particular, we will take the Higgs potential to be (see Fig.~\ref{fig: Higgs potential})
\begin{equation}
\mathcal{L} \supset \mu_h^2 \mathcal{H}^{\dagger} \mathcal{H} - \lambda_h (\mathcal{H}^{\dagger} \mathcal{H})^2 - \frac{(\mathcal{H}^{\dagger} \mathcal{H})^3}{\Lambda_{\mathcal{H}}^2} \label{eq:laghiggs}.
\end{equation}
This potential has the standard electroweak minimum as well as a true minimum at the scale $\vUV \sim \LH$. 
This scale $\LH$ can originate from high energy dynamics such as Grand Unification~\cite{Georgi:1974sy,Pati:1974yy,Dimopoulos:1981zb,Dimopoulos:1981yj}, String Theory, etc.

After inflation, the Standard Model sector can be reheated to a temperature much larger than the scale $\vUV$. During this stage, the Higgs boson receives a thermal correction to its potential that gives the Higgs a large positive mass around the origin and pulls the Higgs field back to the origin. Very quickly, the Higgs decay rate becomes larger than Hubble and it settles around the origin. 
Despite being in the true minimum during inflation, the Higgs ends up in the electroweak minimum. 

Such a scenario is interesting as it provides an opportunity to directly study the Higgs vacuum structure at extremely high energy scales using non-Gaussianities.  
The most pronounced effect due to non-inflaton particles during inflation originates from particles whose masses are close to the Hubble rate.  
The SM fermions, with masses ranging from $y_e \vUV$ to $y_t \vUV$, provide a natural comb that spans more than five orders of magnitude. Some of these fermions will have masses close to the Hubble scale during inflation, leading to observable signatures in the cosmological collider physics program.

Cosmological collider physics provides a new window into the physics surrounding inflation~\cite{Chen:2009zp,Baumann:2011nk,Arkani-Hamed:2015bza,Lee:2016vti,Chen:2016hrz,Meerburg:2016zdz,Kumar:2017ecc}. 
Measurements of the non-analytical pieces of the inflaton three point function can provide information about new particles with masses that are close to the Hubble scale. 
The signal strength depends on both the mass of the new particle as well as its coupling to the inflaton. 
In this paper, we consider the lowest dimensional operator coupling a shift symmetric inflaton with the SM,
\begin{equation}
\mathcal{L} \supset - \frac{c_{f_i} \partial_\mu\phi \, \overline{f_i} \gamma^\mu \gamma^5 f_i}{\Lf} + \cdots , \label{eq:lagfermion}
\end{equation}
where $\phi$ is the inflaton and $f_i$ are the SM fermions
\footnote{We will postpone discussions about the $\phi F \widetilde F$ couplings of the inflaton as they can naturally be a loop factor smaller than the fermion ones with their own distinct phenomenology~\cite{Anber:2009ua}.}.  
This coupling acts like a chemical potential for the broken axial symmetry.  
Thus, it is not surprising that this term can help with particle production~\cite{Chen:2018xck,Adshead:2018oaa}. 
We will work in the framework of effective field theory of inflation~\cite{Weinberg:2008hq,Cheung:2007st} and we will not specify an inflaton model.

Such a coupling breaks Lorentz symmetry for non-zero $\dot \phi$, modifies the fermion dispersion during inflation and leads to particle production during inflation with momentum as large as $\lambda_i = \frac{c_{f_i} \dot{\phi}} {\Lf}$, which can be much larger than Hubble. 
This greatly enhances the number density of fermions produced during inflation and boosts the signal strength in cosmological collider physics, leading to an $\fNL$ that can be as large as
\begin{equation}
\fNL \simeq \Pz^{-1/2} \left(\frac{c_{f_i} m_{f_i}}{\Lf}\right)^3 \lt_i^2 \exp \left[-\frac{\pi m_{f_i}^2}{\lambda_i H}\right] \lesssim 100,
\end{equation}
in the squeezed limit, where $\Pz \simeq 2 \times 10^{-9}$ is the dimensionless power spectrum of curvature perturbations and $\lt_i = \lambda_i/H$ can be as large as $\mathcal{O}(60)$. 
 
The paper is organized as follows. 
In Section~\ref{sec: higgstory}, we discuss in detail the Higgs dynamics during and after inflation. 
In Section~\ref{sec: ccsignature}, we present both the calculation of the non-Gaussianity and a way to estimate the size of the signal. 
In Section~\ref{sec: remarks}, we discuss the future prospect of the measurement of such a signal and the implications for physics beyond the SM.
In the Appendices we collect most of the technical details and further crosschecks.
Appendix~\ref{app: App_NG} contains a detailed exposition of the calculation for the non-Gaussian squeezed bispectrum. Appendix~\ref{app:higgs} discusses some of the details of the Higgs dynamics during inflation as a result of direct Higgs couplings with the curvature and inflaton. 
Appendix~\ref{app:inflaton} discusses the back-reactions on the inflaton dynamics.

\section{Higgs field dynamics in the early universe}
\label{sec: higgstory}

\begin{figure}[h!] \centering
$\hbox{ \convertMPtoPDF{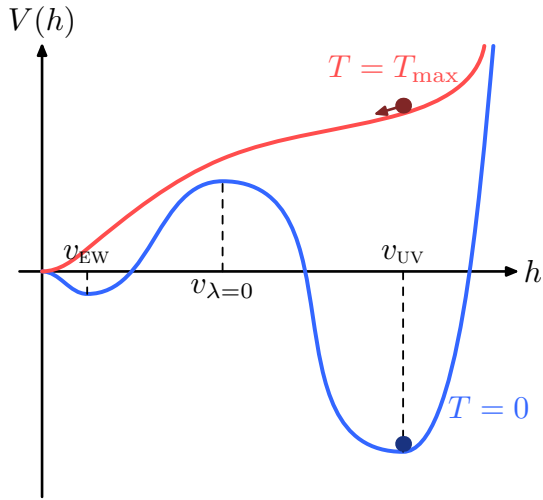}{1.}{1.} }$
\caption{Higgs potential at zero temperature (blue line) and at high temperature during the reheating phase (red line).
The Higgs field sits in the high energy minimum $\vUV$ during inflation, and then returns back to the electroweak vacuum during the thermal phase of reheating, when thermal corrections to the Higgs potential lift the minimum $\vUV$.
In reality, the (free) energy of the Higgs decreases for $|h|<T$, and the thermal potential can be fit by a positive quadratic times exponential term, plus a negative offset that we did not show explicitly in this figure for better presentation.
}
\label{fig: Higgs potential}
\end{figure}

It is well known that, if we extrapolate the running of the SM parameters up to high energies, the Higgs quartic coupling turns negative around the scale $v_{\lambda=0} \sim 10^{11} \GeV$, so that the minimum we live in right now is metastable. 
Beyond the scale of $v_{\lambda=0}$, new physics can come in and save the theory from a runaway direction and create a new minimum of the Higgs potential at some scale $\vUV$. 
The recent upper limit on the tensor to scalar ratio $r<0.06$ \cite{Akrami:2018odb} implies that
\begin{equation}
H<6\cdot 10^{13} \GeV \, , \qquad \text{(upper limit on $r$)}
\label{eq: bound H}
\end{equation}
so that $H$ can still be much larger than the scale $v_{\lambda=0}$. 
During inflation, the Higgs field background undergoes a random walk with kicks $\sim H/(2\pi)$ and could possibly have reached the true minimum at a very large vacuum expectation value (vev) $\vUV$ for the Higgs field. 
The true minimum has a large negative vacuum energy, and the corresponding anti-de Sitter region would expand at the speed of light after the end of inflation \cite{Espinosa:2007qp,Kobakhidze:2013tn,Kobakhidze:2014xda,Fairbairn:2014zia,Enqvist:2014bua,Hook:2014uia,Kamada:2014ufa,Shkerin:2015exa,Kearney:2015vba,East:2016anr,Herranen:2014cua}. 
If inflation occurred at high energy scales, the fact that our observable Universe lies in the electroweak vacuum would seem a very extreme accident and would beg for an explanation. 
In this Section, we describe a scenario in which the Higgs field sits in the true minimum $\vUV$ during inflation and settles back down to the electroweak minimum after reheating. 
We sketch the Higgs potential during and after inflation in Fig.~\ref{fig: Higgs potential}, and we describe in more detail this scenario in the rest of this Section.

\subsection{The Higgs potential during inflation}\label{sec:potentialinflation}

There are a few assumptions about the Higgs potential that need to be satisfied in order for us to observe today the signature of a high energy vacuum. 
For simplicity, we assume that the new minimum for the Higgs field is generated by higher dimensional operators in its potential, suppressed by a cutoff scale $\LH$.
We write then the following Lagrangian for the Higgs field:
\begin{equation}
\mathcal{L}_\text{Higgs} = (\partial_\mu \mathcal H)^\dagger \partial^\mu \mathcal H - V(\mathcal H)\,, \qquad
V(\mathcal H)= -\mu_h^2 \mathcal{H}^{\dagger} \mathcal{H} + \lambda_h (\mathcal{H}^{\dagger} \mathcal{H})^2 + \frac{1}{\LH^2} (\mathcal{H}^{\dagger} \mathcal{H})^3 \,.
\label{eq: Lagrangian Higgs}
\end{equation}

Let us write the Higgs doublet in the unitary gauge as $\mathcal H = (0,\, \tfrac{v+h}{\sqrt 2})^\text T$. 
The mass term of the Higgs potential is irrelevant, being $\mu_h^2\sim \mathcal O(100)\GeV$ and its RG flow negligible.
We assume that $\lambda_h$ turns negative at high energies, so that the potential in Eq.~\eqref{eq: Lagrangian Higgs} has a true vacuum of the Higgs potential at
\begin{equation}
\vUV= \sqrt{\tfrac 43\lUV}\LH\,, \quad \lUV\equiv - \lambda_h(\vUV)>0 \,,
\label{eq: vUV}
\end{equation}
where for the central measured values of $m_t$ and $\alpha_s$ we have $\lUV\sim \mathcal O(0.01)$. 
The uncertainties on the RG evolution of the quartic Higgs coupling mainly come from the uncertainty on the top quark mass, and at subleading order on the strong coupling constant and the Higgs mass \cite{Franciolini:2018ebs}. In this section,
we assume that the RG running of gauge, Yukawa and Higgs quartic ($\lambda_h$) couplings is not affected by new physics between the weak scale and $\LH$. 
We postpone discussion of the effect of a Higgs coupling to curvature and the inflaton to Appendix~\ref{app:higgs}. 
These corrections can increase the Higgs field value during inflation and can lead to interesting observable effects~\cite{Hook:2019}.

Depending on the Hubble rate, which we assume to be comparable to the current bound \eqref{eq: bound H}, and the number of $e$-folds of inflation, the Higgs field can easily overcome the barrier under the effect of quantum fluctuations, and reach its true minimum $\vUV$ within a few $e$-folds. 
We assume that $\langle h\rangle \sim \vUV$ throughout the $\sim 60$ $e$-folds of inflation that we can potentially observe today.

\paragraph{Quantum fluctuations of $\langle h \rangle$.} The Higgs field can fluctuate around $\vUV$ during inflation by steps of order $H/2\pi$, which can lead to a fluctuation of the fermion masses during inflation. 
In this paper, we will restrict our analysis to the case where these Higgs fluctuations are negligible such that the statistical errors are reduced and the predictions are much simpler. 
This requires that the Higgs mass squared at $\vUV$ is greater than $9 H^2/4$.  
In this case, fluctuations are exponentially suppressed. 
This leads to a constraint
\begin{equation}
\frac 94 H^2 < V''(\vUV) = \frac 83 \lUV^2 \LH^2 
\quad \Rightarrow \quad
\vUV > \sqrt{\frac{9}{8 \lUV}} H 
\qquad \text{(no fluctuations at $\vUV$)}
\label{eq: no fluct v}
\end{equation}
If the Higgs is subject to quantum fluctuations, but the spread in field values induced during a number $N\gtrsim \mathcal O(10)$ e-folds is not larger than $\vUV$, then the spatial variations of $\langle h\rangle$ on the scales probed by present day experiments would still be small.
The non-Gaussianity estimates that follow are still valid, with some small quantitative differences\footnote{This scenario could imply exciting distinctive signatures, like spatial variations of $\fNL$.}, as long as
\begin{equation}
\vUV \gtrsim H \frac{\sqrt N}{2\pi} \overset{N\sim \mathcal O(10)}\simeq H
\qquad \text{(small fluctuations around $\vUV$)}
\label{eq: small fluct v}
\end{equation}

\paragraph{No alterations of the inflationary dynamics.} In order for the inflationary dynamics not to be significantly affected by the negative Higgs energy density when $\langle h\rangle =\vUV$, the sum of $V_h(\vUV)$ and $V_\phi=3H^2\MP^2$ must be positive:
\begin{equation}
|V_h(\vUV)| = \left|- \frac{4}{27}\lUV^3\LH^4 \right| <
 3H^2 \MP^2 
 \quad \Rightarrow \quad \vUV< \frac{\sqrt{6\MP H}}{\lUV^{1/4}} \,.
 \qquad \text{($V_\phi>|V_h|$)}
 \label{eq: constraint no AdS}
\end{equation}
This constraint, for $\lUV\sim 0.01$, turns out to be weaker than the requirement \eqref{eq: constr thermal rescue} that the temperature is high enough to bring the Higgs vev back to the origin after inflation ends.

\begin{figure}[h!]
\centering
\includegraphics[width=0.6\textwidth]{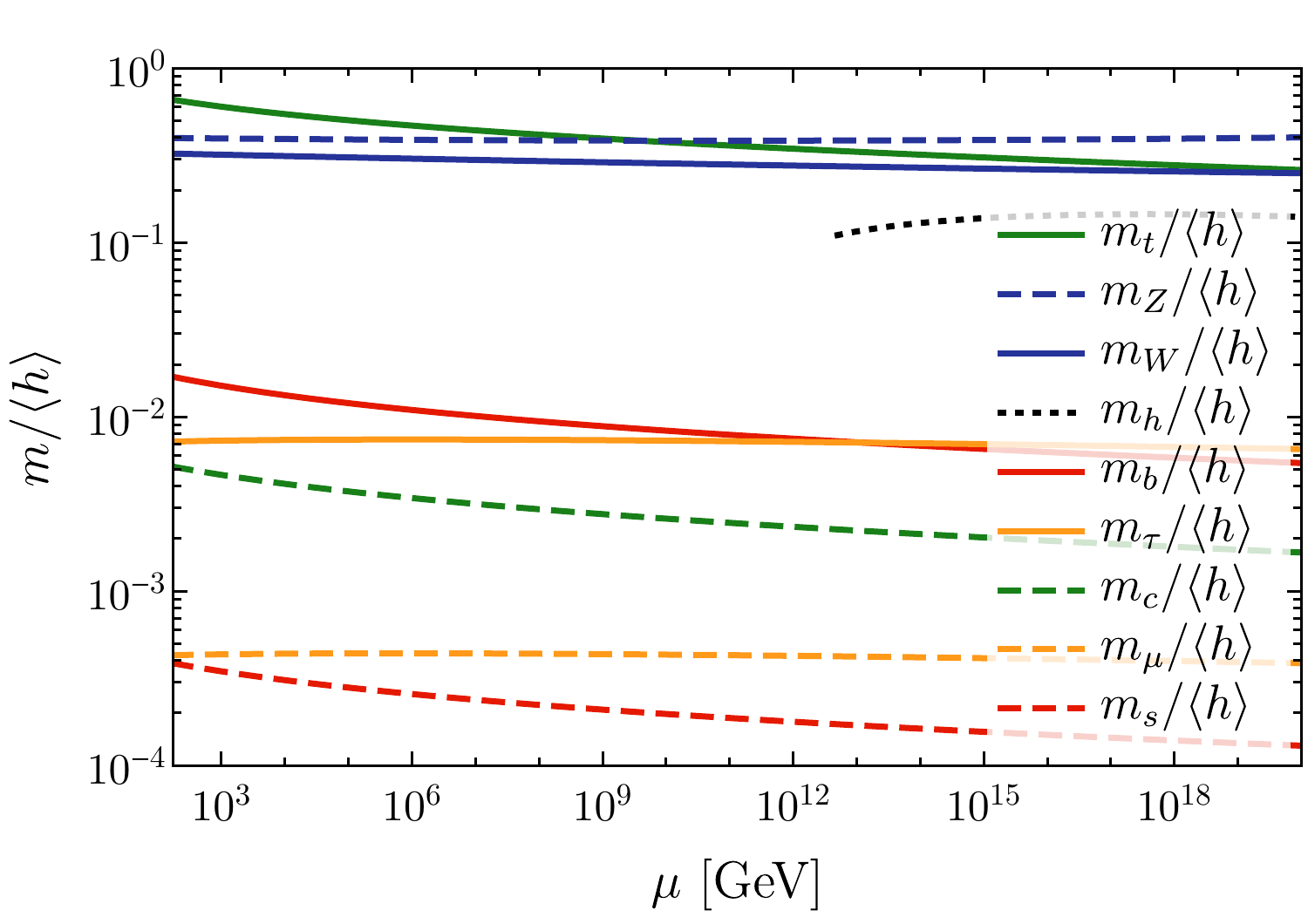}
\caption{RG flow of the coefficients of the masses of the SM particles in terms of the Higgs vev. 
The width of the lines in this plot is the larger than the current experimental uncertainty on the SM parameters at the weak scale.
}
\label{fig: running SM}
\end{figure}

\paragraph{Lifting of the SM mass spectrum.}
Once the Higgs is in the UV minimum, the spectrum of all SM particles during inflation are solely determined by a single parameter $\LH$.
In terms of the Higgs vev $\vUV$, we collect here the masses of the SM particles: 
\begin{equation}
\begin{aligned}
m_{\Psi_i} =& \frac{1}{\sqrt 2} y_i \, \vUV \\
m_{h} =& \sqrt{2\lUV} \,\vUV \\
m_{W} =& \frac 12 g_2 \,\vUV\\
m_{Z} =& \frac 12 \sqrt{g_1^2+ g_2^2} \, \vUV
\end{aligned}
\label{eq: masses SM}
\end{equation}
We show in Fig.~\ref{fig: running SM} the running of the coefficients appearing in the masses in Eq.~\eqref{eq: masses SM}. 
\newline
In Fig.~\ref{fig: plane v H} we show the lines in the parameter space corresponding to $H=m_f$ for the SM fermions.

The wealth of massive particles due to the UV Higgs minimum spans five orders of magnitude.
Therefore, it is very likely that one or two of them will happen to have a mass close to the Hubble scale. 
If we detected the signature of the presence of two or three fermions with mass ratios resembling those of the Yukawa couplings, it would be a very strong indication for the existence of a new Higgs minimum at high scales. 
We show how to estimate and calculate the amount of non-Gaussianity that can arise due to these new fermions in Sec.~\ref{sec: ccsignature}.

\subsection{Higgs potential during reheating}

The Higgs field will need to find its way back to the symmetry preserving point $h=0$ after inflation. This happens if the universe reheats to high enough temperatures $\TRH\gtrsim \vUV$, where thermal corrections to the Higgs potential can bring the Higgs vev from $\vUV$ back to the origin.
Reheating generates a thermal bath of SM particles which contribute to the Higgs potential with a thermal mass  
\cite{Espinosa:2015qea,Ema:2016kpf,Kohri:2016wof,Enqvist:2016mqj,Postma:2017hbk,Ema:2017loe,Joti:2017fwe,Rajantie:2017ajw} 
\begin{equation}
V_T(h) \simeq \frac 12 \kappa T^2\,h^2 e^{-h/(2\pi T)}\,, \quad \kappa \simeq 0.12 \,.
\end{equation}
This contribution pushes the peak of the barrier in the Higgs potential to values equal to roughly twice the temperature. With the addition of the thermal contribution to the potential, the Higgs field rolls back and forth in the potential during the reheating phase and decays into SM matter.

The requirement of the rescue of the Higgs field can be converted into a bound on the maximum temperature reached during reheating.
Assuming for simplicity instantaneous reheating, then all the inflaton energy density is completely converted at the end of inflation into thermal radiation fluid at a reheating temperature $\TRH$ given by $(\pi^2/30) g_* \TRH^4 = 3H^2 \MP^2$, $\MP$ being the reduced Planck mass and $g_*=106.75$ the number of SM relativistic degrees of freedom at early times.
From Eq.~\eqref{eq: vUV} we get 
\begin{equation}
\begin{aligned}
\LH & = \frac{\vUV}{\sqrt{\tfrac 43 \lUV}} 
      \lesssim \frac{\TRH}{\sqrt{\lUV}} = \\
      & = \frac{(6.5 \cdot 10^{15}\GeV)}{\sqrt{\lUV}} \left(\frac{H}{6\cdot 10^{13}\GeV} \right)^{1/2}
\end{aligned}
\qquad \text{(Higgs thermally rescued)}
\label{eq: constr thermal rescue}
\end{equation}
This is the condition that ensures that the Higgs is rescued by thermal corrections during reheating.
By plugging $\lUV\sim 0.01$, we get that at most $\LH\lesssim 10^{16} \GeV$. 
Given that the reheating temperature $\TRH$ can be larger than the typical scale $\LH$ of higher dimensional operators, this calculation is not technically under control. 
However, ultraviolet completions of the theory might not change the results significantly.

As explained in more detail in App.~\ref{app:higgs}, the magnitude of the oscillations of the Higgs field decreases rapidly and the Higgs very quickly relaxes to the origin.
Therefore the Higgs field eventually lays at the origin, provided that the initial condition \eqref{eq: constr thermal rescue} is satisfied. 
We postpone more detailed discussions to appendix~\ref{app:higgsreheat}.

\subsection{Summary of the viable parameter space}
\begin{figure}[h!]
\centering
\includegraphics[width=0.8\textwidth]{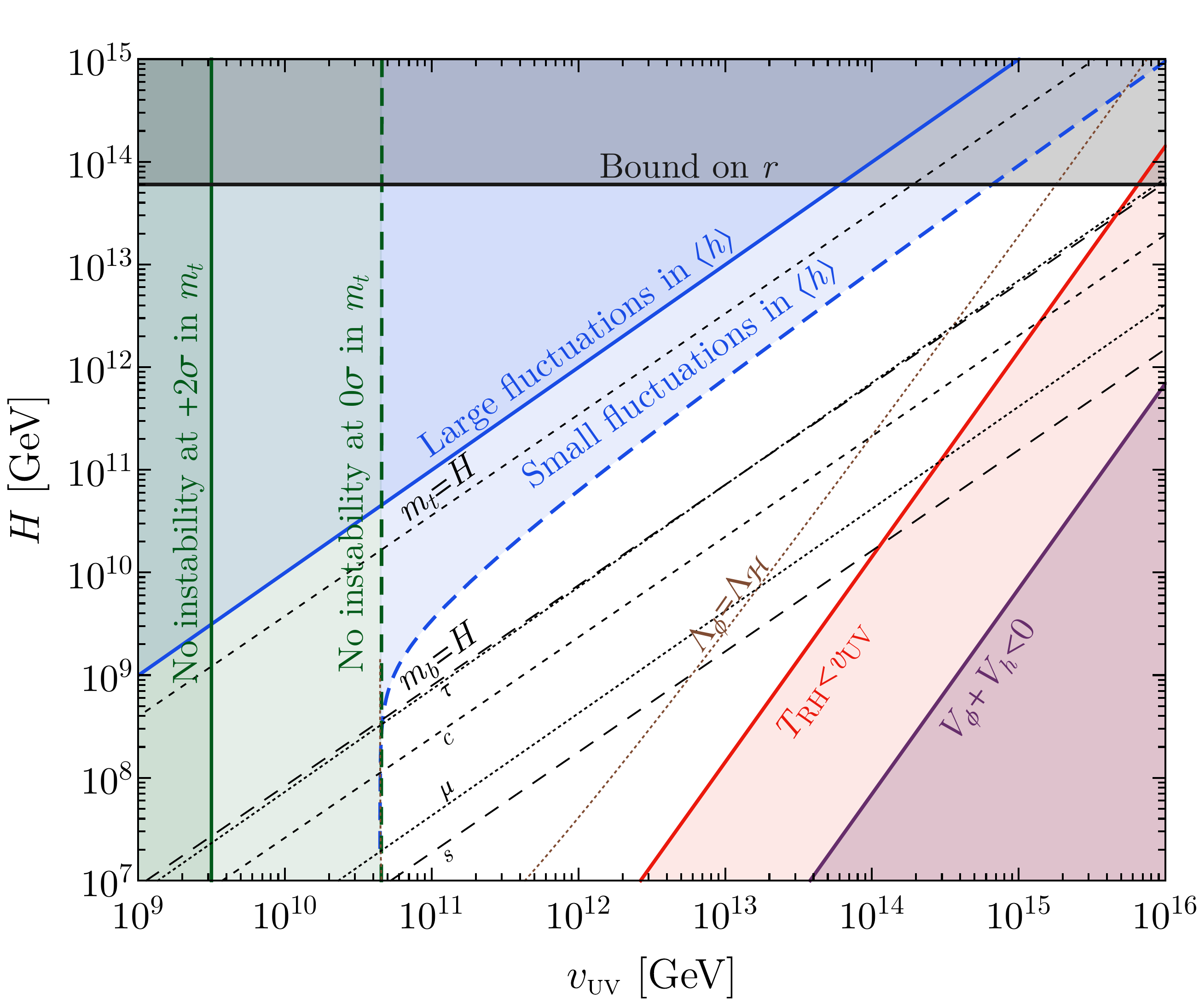}
\caption{Viable parameter space in the plane $(\vUV,H)$.
The shaded regions are excluded due to following constraints, listed also in the text: 
$H$ allowed by the bound on $r$ (gray), 
negligible quantum fluctuations at $h=\vUV$ (blue),
Higgs quartic turning negative within the SM (green),
high enough $\TRH$ to rescue the Higgs (red),
Higgs energy density smaller than the inflaton one (purple).
The thin black lines show where $m_f=H$ for the SM fermions.
The thin dotted brown line corresponds to the case in which the energy scale $\Lphi$ of the inflaton is equal to $\LH$. 
}
\label{fig: plane v H}
\end{figure}

We show in Fig.~\ref{fig: plane v H} the constraints in the plane $(\vUV, H)$ arising from the following considerations:
\begin{enumerate}
\item \textit{gray line}: upper bound on $H$ from the constraint on $r$, see Eq.~\eqref{eq: bound H};
\item \textit{blue lines, dashed}: no quantum fluctuations of the Higgs at $\vUV$, see Eq.~\eqref{eq: no fluct v}, plugging the running of $\lUV$ for the central measured SM values; \textit{solid}: small quantum fluctuations, as in Eq.~\eqref{eq: small fluct v};
\item \textit{green lines}: presence of the instability (i.~e.\ $\lambda (\vUV)<0$) within the SM, for the central measured values in $\alpha_s$ and $m_h$ and $0\sigma$ or $+2\sigma$ deviations in $m_t$  (for the reference values, see \cite{Franciolini:2018ebs});
\item \textit{red line}: energy scale during inflation giving a high enough temperature to rescue the Higgs after inflation, assuming instantaneous reheating (see Eq.~\eqref{eq: constr thermal rescue});
\item \textit{purple line}: Higgs negative energy density never overcoming the inflaton energy density (see Eq.~\eqref{eq: constraint no AdS}); this constraint is weaker than the previous one.
\end{enumerate}
In order to highlight the most interesting regions for the signature we discuss, we show with thin black lines where the Hubble rate equals the mass of a SM fermion. 
We also show with a thin brown line where the inflaton energy scale $\Lphi$ during inflation, defined as $3H^2\MP^2 = \Lphi^4$, is equal to $\LH = \sqrt{3/4\lUV}\,\vUV$.
\newline
As we will see, our signal is generically amplified for larger values of $H$, so that the most promising region is the one around $H\sim 10^{13}\GeV$ and $\vUV\sim 10^{14}-10^{15}$ GeV (corresponding to $\LH\sim 10^{15}-10^{16}\GeV$), for which $H$ is close to the masses of the $b$ and $\tau$ fermions.

\section{Cosmological collider signature}
\label{sec: ccsignature}


In this Section, we present the calculation, together with a more physical interpretation, of the non-Gaussianity coming from SM fermions coupled to the inflaton. 
We focus on the main steps of the calculation and move most of the details to Appendix~\ref{app: App_NG}. 
In Section~\ref{sec:otheroperator}, we discuss the effect of some other operators coupling the inflaton to the SM. 
Readers who are mainly interested in the implications of the effect can skip Section~\ref{sec: fNL calculation}.

\subsection{How to estimate $\fNL$}\label{sec:estimate}
\label{sec: fNL estimate}

\begin{figure}[h!] \centering
\includegraphics[width=0.45\textwidth]{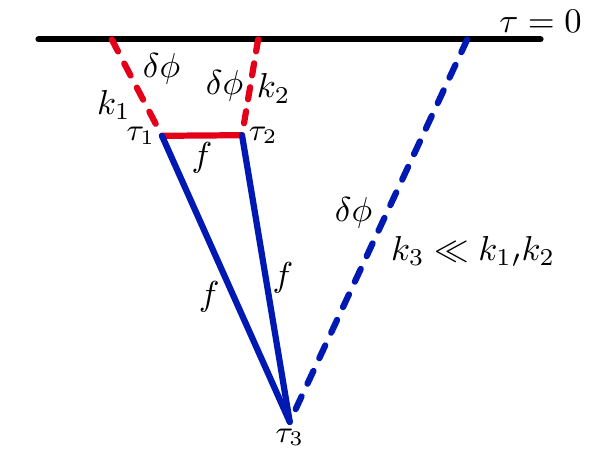}
\caption{
Feynman diagram for the contribution to the 3-point function of the inflaton from a loop of SM fermions.
Two SM fermions $f$ are produced at an early time $\tau_3$ through an interaction with a soft inflaton fluctuation $\delta\phi$, and annihilate at later times $\tau_1$, $\tau_2$ producing two hard inflaton legs with $k_1,k_2\gg k_3$.
The time propagation of the fermions $f$ is at the origin of the non-analytic term in the bispectrum of the inflaton.
}
\label{fig:dinf}
\end{figure}

In this subsection, we briefly outline how one estimates non-Gaussianity in the context of cosmological collider physics. 
As with many things, a good starting point is the definition.   
Throughout this subsection, we work in units where $H = 1$. 
In spatially flat gauge ($\mathcal{R} = \zeta = - \delta \phi/\dot \phi$), the two point function is
\begin{equation}
\Big\langle \zeta(k) \zeta(-k) \Big\rangle' = \frac{2\pi^2}{k^3}\Pz = \frac{1}{\dot \phi^2} \frac{1}{2 k^3} \,,
\end{equation}
where we denote by a dot the derivatives with respect to cosmic time. 
We also denote by $\tau$ the conformal time defined as usual by $\mathrm d\tau = \mathrm dt/a$.
We adopt the primed notation, defined as
\bea
\Big\langle \delta \phi(k_1) \cdots \delta \phi(k_n) \Big\rangle = (2 \pi)^3 \delta\Big(\sum_i k_i\Big)  \Big\langle \delta \phi(k_1) \cdots \delta \phi(k_n) \Big\rangle' .
\eea
One of the dimensionless functions which characterizes non-Gaussianities is the dimensionless shape $S(k_1, k_2, k_3)$,
\bea
\langle \zeta(k_1) \zeta(k_2) \zeta(k_3) \rangle' =  \frac{(2 \pi)^4 \Pz^2}{k_1^2 k_2^2 k_3^2} S(k_1, k_2, k_3).
\eea
We will be interested in the non-analytic part of the squeezed limit ($k_1 \sim k_2 \gg k_3$)
\bea
S = S(k_1,k_2,k_3)^{\text{non-analytic}} \Big|_{k_3 \ll k_1 \sim k_2} .
\eea
We want to estimate the diagram shown in Fig.~\ref{fig:dinf}. 
\newline
The first point to assess is what is the exponential suppression.  Particle production occurs when the adiabatic approximation fails.  
The exponential suppression associated with adiabatic processes is
\bea
e^{- \frac{\omega^2}{\dot \omega}} \sim e^{- \frac{\omega^2}{\tau \, \mathrm d\omega/ \mathrm d\tau}}.
\eea
The time scale $\tau$ and the exponential suppression can be found by minimizing the exponential suppression.  
In our case, the leading terms in the dispersion relations for the fermions are
\bea
\omega^2 = (k \tau \pm \lambda)^2 + m^2
\label{eq: dispersion relation}
\eea
Taking the large $\lambda$ limit and minimizing, we get that the exponential suppression is $\omega^2/\dot \omega \sim m^2/\lambda$ which occurs when $k \tau \sim  \lambda$ with a width of order $m$.  
Our first goal is to consider the large $\lambda$ limit and obtain the exponential suppression in this limit. 
This is clearly
\bea
S \sim e^{-\frac{m^2}{\lambda}} \Big|_{1 \ll m^2/\lambda \ll \lambda}
\label{eq: fNL exp suppr}
\eea
The next limit we wish to consider is the small $m$ limit.  In this limit
\bea
n \sim k^2 \delta k  \sim m \lambda^2 \Big|_{m \ll 1}
\eea
The next thing to estimate is the momentum dependence of $S$.  There is no quick trick we know of to directly obtain the analytic pieces of the momentum dependence, so we focus on the non-analytic contributions. The non-analytic piece comes from the propagators of the fermions.  
Of the three propagators, one of them has a large momentum running through it and thus is insensitive to the effects of Hubble and can be ignored. 
The two remaining propagators each contain a factor of $e^{\pm i \omega t} \sim \tau^{\pm i\omega} \sim k^{\pm i\omega}$.  
Thus we find that the scaling of the non-analytic piece is
\bea
S \sim m \lambda^2 \left ( \frac{k_3}{k_1} \right )^{2 i \lambda} 
\eea
We are not aware of a simple way to estimate the non-imaginary part of the exponent of $k_3/k_1$.

The last factors associated with the non-Gaussianities are the coupling constants.
There are three insertions of the inflaton so there is a factor of $1/\Lf^3$.  
Finally, by doing a field redefinition, derivative interactions with the current can be shown to be proportional to the mass. 
The hard propagator can be effectively integrated out giving only two mass insertions.  
Thus there is an additional factor of $m^2$ in the small $m$ limit.  
We arrive at our final estimate of the non-Gaussianity
\bea
S \sim m^2 \frac{1}{\Lf^3} m \lambda^2 \left ( \frac{k_3}{k_1} \right )^{ 2 i \lambda} \Bigg|_{\lambda \gg 1,\, m\to 0}.
\label{eq: fNL estimate}
\eea
The two estimates \eqref{eq: fNL exp suppr} and \eqref{eq: fNL estimate} of the non-Gaussianity, valid respectively in the limits of large $m^2/\lambda$ and large $\lambda$, small $m$, have the scaling found in an explicit calculation.
 
\subsection{Outline of the calculation of $\fNL$}
\label{sec: fNL calculation}

\begin{figure}[h!] \centering
\includegraphics[width=0.7\textwidth]{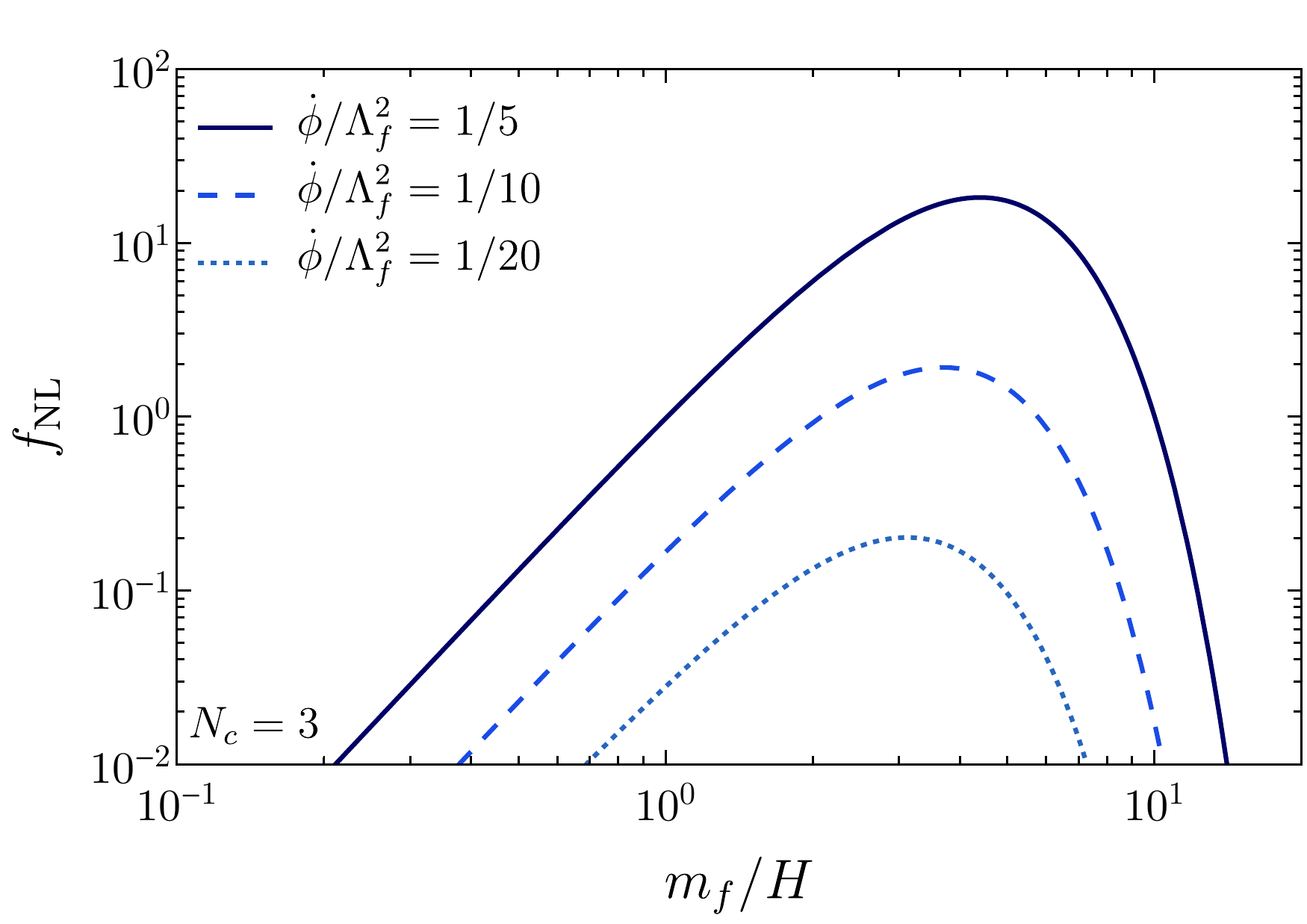}
\caption{The signal strength due to a single Dirac fermion as a function of the fermion mass in Hubble units. 
The solid, dashed and dotted blue lines show $\fNL$ for different values of $\dot \phi/\Lf^2 \leq 1$. 
The signal strength increases as $m^3$ in the small mass limit and decreases exponentially when $(m/H)^2$ becomes larger than $\frac{\dot{\phi}}{\Lambda_f H}$.}
\label{fig:fnlml}
\end{figure}

In this subsection, we present the main steps of the computation which leads to results presented in Figures~\ref{fig:fnlml}, \ref{fig:signalf} and \ref{fig:signaltop}. 
The main Feynman diagram that contributes to the three point correlation function is shown in Fig.~\ref{fig:dinf} where the dashed lines represent the inflaton perturbation $\delta \phi$, the solid lines represent a SM fermion $f$ and the vertex comes from the interaction of Eq.~\eqref{eq:lagfermion} (for detailed Feynman rules, see Appendix~\ref{app: App_NG}),
\begin{equation}
\mathcal{L} \supset - \frac{c_{f_i} \partial_{\mu}{\phi} \overline{f_i} \gamma^{\mu} \gamma^5 f_i}{\Lf} \,. \label{eq:interaction}
\end{equation}
This coupling between the inflaton and the fermions leads not only to an interaction vertex between $\delta\phi$ and $f$ but also to a correction to the dispersion relation of the SM fermions when the inflaton slow-roll spontaneously breaks Lorentz symmetry:
\begin{equation}
\omega^2 = \left(k \mp \lambda_i\right)^2 + m_i^2
\label{eq: dispersion relation flat}
\end{equation}
in flat space, where $k =|\vec{k}|$, $\lambda_i = \frac{c_{f_i} \dot{\phi}} {\Lf}$ and $\pm$ marks states with different helicity. 
In the following, unless stated explicitly, we consider a single fermion with $c_{f_i}=1$, $\lambda = \lambda_i$ and $m=m_i$. 
The correction to the dispersion relation leads a modification of the fermion mode functions $u_s$ and $v_s$. 
The solution reads (for a complete list of the mode functions, see Eq.~\eqref{eq: fermion Whittaker})
\begin{equation}
u_+ (k\tau)= \frac{\mt e^{\pi \lt/2}}{\sqrt{-2 k \tau}} W_{\widetilde \kappa, i \mut} (2 i k \tau)\,,
\end{equation}
where $\pm$ are again the helicity indices, $W_{\kappa,\mu} (z)$ is the Whittaker function, and we define dimensionless parameters with a tilde:
\begin{equation}
\widetilde{\kappa} = -\frac{1}{2} - i\lt, \quad \mt = m/H, \quad \lt = \lambda/H, \quad \mut = \sqrt{\mt^2+\lt^2}.
\end{equation}
At late times ($-k \tau \ll 1$), the mode function $u_+$ has the following dependence on $k\tau$ (see Appendix~\ref{app: fermion mode functions} for more details):
\begin{equation}
u_+ (k\tau) \simeq e^{- i \pi/4}\mt e^{\pi \lt/2}\left[\frac{e^{\pi \mut/2} \Gamma(-2 i \mut)}{\Gamma(1+ i\lt -i \mut)} (- 2 k \tau)^{i \mut} + (\mut\rightarrow -\mut)\right].
\end{equation}
This is to be compared with the late-time limit of a particle with an ordinary dispersion relation where one gets the dependence of $(- 2 k \tau)^{i \mt}$ in the large mass limit instead of $(- 2 k \tau)^{i \mut}$. This can be understood as a result of the abnormal ``redshifting'' of the fermions during inflation in our case, where as the momentum of the fermion decreases, the frequency quickly increases from $\mathcal{O}(m)$ to $\mathcal{O}(\sqrt{m^2+\lambda^2})$. This oscillation frequency turns into the frequency of oscillation of $k_3/k_1$ in the final result. 
Such a late time expansion is clearly not valid around the dominant time of particle production when $-k\tau\sim \lt\gg 1$, which leads to a numerical difference between our result and that of \cite{Chen:2018xck} (see Appendix~\ref{sec: time integrals} for a mathematical treatment of the discrepancy).

The physical process that happens during inflation is shown in Fig.~\ref{fig:dinf}. 
An inflaton perturbation with a soft momentum $k_3$ splits at some early time $\tau_3$ into two fermions both with momentum $k\sim \lambda$ and frequency $\omega \sim m$.%
\footnote{
More precisely, the physical picture is that the fermions get produced at some time before $\tau_3$ and shortly after annihilate at $\tau_3$, before their momentum gets significantly reshifted. 
For simplicity, we identify $\tau_3$ with the time of fermion creation.} 
Then the fermions redshift and annihilate back into inflaton perturbations at much later times $\tau_1$, $\tau_2$. 
The 3-point function of the inflaton perturbation $\delta \phi$ generated by this process is
\begin{multline}
\left\langle\delta \phi (\vec k_1) \delta \phi ( \vec k_2) \delta \phi (\vec k_3)\right\rangle = \\
\sum_{\oa,\ob,\oc =\pm 1} \oa\ob\oc \left(\frac{i}{\Lambda}\right)^3 \iiint_{-\infty}^0 \mathrm{d} \tau_1 \mathrm{d} \tau_2 \mathrm{d} \tau_3 F_{\mu\oa}(\vec k_1,\tau_1) F_{\nu\ob}(\vec k_2,\tau_2) F_{\rho\oc}(\vec k_3,\tau_3)
\int \frac{\mathrm d^3q}{(2\pi)^3} \mathscr T^{\mu\nu\rho}_{\oa\ob\oc},\label{eq:amplitude1}
\end{multline}
where $F_{\mu\oa}(\vec k_1,\tau_1)$ (see Eq.~\eqref{eq:Ffunction}) comes from the external leg of the inflaton perturbation $\delta\phi (\vec k_1)$. 
The trace $\mathscr T^{\mu\nu\rho}_{\oa\ob\oc}$ originates from the fermion loop:
\begin{equation}
\begin{aligned}
\mathscr T^{\mu\nu\rho}_{\oa\ob\oc} =& -\text{tr}\left[
\overline \sigma^{\mu\dot\alpha \alpha} D_{\oa \ob \alpha \dot \beta}(p_{12},\tau_1,\tau_2)
\overline \sigma^{\nu\dot\beta \beta} D_{\ob \oc \beta \dot \gamma}(p_{23},\tau_2,\tau_3)
\overline \sigma^{\rho\dot\gamma \gamma} D_{\oc \oa \gamma \dot \alpha}(p_{31},\tau_3,\tau_1)
\right] \\
&  -\text{tr}\left[
\overline \sigma^{\mu\dot\alpha \alpha} D_{\oa \oc \alpha \dot \gamma}(-p_{31},\tau_1,\tau_3)
\overline \sigma^{\rho\dot\gamma \gamma} D_{\oc \ob \gamma \dot \beta}(-p_{23},\tau_3,\tau_2)
\overline \sigma^{\nu\dot\beta \beta} D_{\ob \oa \beta \dot \alpha}(-p_{12},\tau_2,\tau_1)
\right]
\end{aligned}\label{eq:trace1}
\end{equation}
where $D_{\oa \ob \alpha \dot \beta}(p,\tau_1,\tau_2)$ are the propagators of the fermions. 
The previous discussion motivates us to split up the fermion propagators into the mode functions $u$ and $v$, where $u(k_3 \tau_3)$ and $v(k_3 \tau_3)$ can be expanded in the large $\lambda$ limit while the functions $u(k_3 \tau_1)$ and $v(k_3 \tau_1)$ can be expanded in the late time limit ($-k_3 \tau_1\ll 1$). 
This allows us to turn the trace in Eq.~\eqref{eq:amplitude1} into functions over which we can perform the integral over the times $\tau_1$, $\tau_2$ and $\tau_3$ (see Appendix~\ref{sec: time integrals} for more details). 
These integrals are, as physically motivated, dominated by regions where $-k_i \tau_i \sim \tilde{\lambda}$. This suggests that our results are only valid for momentum ratios 
\begin{equation}
k_3/k_1 = \frac{k_3 \tau_1}{k_1 \tau_1} \lesssim 1/\lt.
\end{equation}
This momentum ratio can be understood as the ratio of energies of a physical process where the energy of the fermions is $\mathcal{O}(m)$ to start with (the momentum and energy of the inflaton leg $k_3$ is comparable to the energy of the intermediate fermions when they are produced $\omega(\tau_{3}) \sim m$), and $\mathcal{O}(\sqrt{\lambda^2+m^2})$ in the late time limit when the two fermions annihilate (the momenta and energies of the inflaton legs $k_1 \sim k_2$ are comparable to the energy of the intermediate fermions when they annihilate $\omega(\tau_{1,2}) \sim \sqrt{\lambda^2+m^2}$)%
\footnote{These discussion should provide an estimate of the leading dependence on $\lambda$ in the large $\lambda$ limit. 
As is clear from the detailed calculation in Appendix~\ref{app: App_NG}, the $m$-dependence in the $H<m<\lambda$ case can be very complicated as it can receive contributions from various sources.}. 
This requirement matches the expectation that the result we obtained in this calculation should not have an enhancement of $\lambda^2$ in the limit where the fermion exchange can be treated as a contact operator. 
To conclude, the result of the full calculation at leading order in the squeezed limit (accounting for two chiralities) is
\begin{equation}
S (k_1, k_2, k_3) \overset{\lambda \gg m }{\overset{k_3\ll k_1\sim k_2}{\simeq}} \fNL^{({\rm clock})} \left(\frac{k_3}{k_1}\right)^{2-2 i \tilde{\mu}} + \cdots
\end{equation}
up to a phase, with
\begin{equation}
f_{\rm NL}^{({\rm clock})}  \approx \frac{N_c}{6\pi}  \Pz^{-1/2} \left(\frac{m}{\Lf}\right)^3 \lt^2  \frac{e^{\pi  \lt } \mut  \Gamma (-i \mut )^2 \Gamma (2 i \mut )^3}{2 \pi  \Gamma (i (\lt+\mut))^3 \Gamma (i (\mut -\lt)+1)}\label{eq:fnl}
\end{equation}
for each SM fermion with color number $N_c$ (we recall that we have set $c_{f_i}=1$, and it can be restored by replacing every occurrence of $1/\Lf$ by $c_{f_i}/\Lf$). 
From now on we refer to $\fNL$ as the amplitude of the clock signal $f_{\rm NL}^{({\rm clock})}$, defined in Eq.~\eqref{eq:fnl} in agreement with what done in \cite{Chen:2018xck}.
In the limit where $m \rightarrow 0$ and $\lambda\rightarrow \infty$, the result scales as $\left(\frac{m}{\Lf}\right)^3 \lambda^2$ at leading order, while in the limit where $H \ll m^2/\lambda \ll \lambda$, there is an exponential suppression from the last factor in Eq.~\eqref{eq:fnl}
\begin{equation}
\exp\left[-\frac{\pi m^2}{\lambda H}\right],
\end{equation} 
both as expected from Sec.~\ref{sec: fNL estimate}. 
The result of the computation is summarized in Fig.~\ref{fig:fnlml}. 
We highlight the dependence of the signal strength on the fermion mass in Hubble units. 
As expected, the signal strength is maximized when the exponent $\pi m^2/\lambda H$ is $\mathcal{O}(1)$, and the size of $\fNL$ can be $\mathcal{O}(10)$ for perturbative coupling.

\subsection{Other operators}
\label{sec:otheroperator}
Other operators coupling the inflaton to SM particles that can potentially lead to observable effects during inflation are studied in \cite{Kumar:2017ecc}. 
Firstly, there can be operators that couple the inflaton to spin-zero marginal operators of the SM in a shift-symmetric manner
\begin{equation}
\mathcal{L} \supset f\left(\frac{(\partial \phi)^2}{\LH^4}\right)\mathcal{O}_{\rm SM}^{(4)}\,,
\end{equation}
where $f(x)$ is a polynomial function of $x$ with order one coefficients. 
Similar couplings between the Higgs and inflaton, in the absence of a large Higgs vev, lead to negligible contributions to $\fNL$, smaller than $\mathcal{O}\left(\frac{{H}^4 \dot \phi^2}{\LH^8}\right)$. 
These effects are unlikely to be observable. 
Inflaton couplings to relevant operators in the SM can potentially lead to much stronger effect. 
However, as we show in more detail with the following examples, the fermion coupling we consider is the only coupling that can lead to large observable effects.
This can be ultimately seen as a consequence of the ``hierarchy problem'' of the Higgs boson.

\paragraph{The case of $\mathcal{H}^{\dagger}\mathcal{H}$} 
An example of such interaction considered in \cite{Kumar:2017ecc} is the operator 
\begin{equation}
\mathcal{O}_{h2} = \frac{c_2 (\partial \phi)^2}{\LH^2}\mathcal{H}^{\dagger}\mathcal{H} \,.
\end{equation}
This operator, similarly to the $\xi_h R \mathcal{H}^{\dagger}\mathcal{H}$ coupling, can be generated by integrating out SM fermions, and leads to a contribution to the Higgs mass during inflation
\begin{equation}
\mu_h^2 \sim \frac{c_2 \dot{\phi}^2}{\LH^2}\,.
\end{equation}
The operator $\mathcal O_{h2}$ can lead to interesting changes to the Higgs potential during inflation (see the companion paper~\cite{Hook:2019} for more details) but is hard to observe: increasing the coupling $c_2$ simultaneously increases the strength of the signal, as well as the Higgs mass, which suppresses exponentially the contribution of a Higgs boson loop to the bispectrum. 
Therefore, in absence of Higgs mass tuning, the signal strength is likely quite small \cite{Kumar:2017ecc}. 
Such a conflict is a direct consequence of the hierarchy problem of the Higgs, which is why it does not affect the signal of the fermions. 

\paragraph{The case of $\mathcal{H}^{\dagger}\mathcal{D} \mathcal{H}$} 
A special case of derivative coupling is the operator 
\begin{equation}
\mathcal O_{h1} = \frac{c_1 (\partial_{\mu} \phi)}{\LH}\mathcal{H}^{\dagger}\mathcal{D}^{\mu}\mathcal{H},
\end{equation}
which couples the inflaton to the Higgs current. 
Such an operator introduces a mixing between the Higgs and the time component $Z^0$ of the $Z$ boson in the form of
\begin{equation}
\frac{{\rm Im}(c_1) \dot{\phi} g_2 \vUV/2}{\LH} h Z^0,
\end{equation} 
where the explicit dependence on $Z^0$ is a sign of broken Lorentz symmetry. 
A large $c_1$ coupling will also lead to significant changes to the UV potential of the Higgs. 
During inflation, the $Z^0$ field will acquire a vacuum expectation value of order $\left(c_1 \frac{\dot{\phi}^2}{\LH^2}\right)^{1/2}$, which in turn leads to a mass and vev of the Higgs of the same order. 
As a result, also the operator $\mathcal O_{h1}$ is not likely to be observed, because the increase in the signal due to a larger $c_1$ is vastly overcome by a severe exponential suppression due to a larger $Z$ mass.

\paragraph{The case of $G\widetilde G$} 
The CP violating coupling between the inflaton and the gauge bosons \begin{equation}
c_G \frac{\phi}{\Lambda_G} G\widetilde G,
\end{equation}
where $G$ stands for a gauge boson of the SM gauge group $SU(3)_C\times SU(2)_L \times U(1)_Y$, is secretly a derivative coupling, and can be generated if there is a gauge anomaly (see Appendix~\ref{app:inflaton}). 
Particle production as a result of this coupling has been studied in depth in the literature \cite{Anber:2009ua,Barnaby:2010vf,Barnaby:2011qe}. 
A large inflaton-gauge boson coupling leads to exponentially growing production of the gauge boson and can strongly affect the inflaton dynamics. 
In our case, if the fermion current the inflaton couples to is anomalous, an inflaton-gauge boson coupling can arise at loop level, with a coupling strength $\frac{\dot{\phi}}{\Lambda_G H} \sim \frac{\alpha_{1,\,2,\,3}}{4\pi} \frac{\dot{\phi}}{\Lf H} \lesssim 1$. 
This means that, in absence of some inherited anomaly, the couplings we write down is unlikely to lead to significant production of the gauge bosons, especially the massless gluons and photons%
\footnote{Such an exponentially growing production is also cut off by the scattering or annihilations of gauge bosons in the SM.}.

\section{Result and implications}
\label{sec: remarks}

\begin{figure}[h!] \centering
\includegraphics[width=0.8\textwidth]{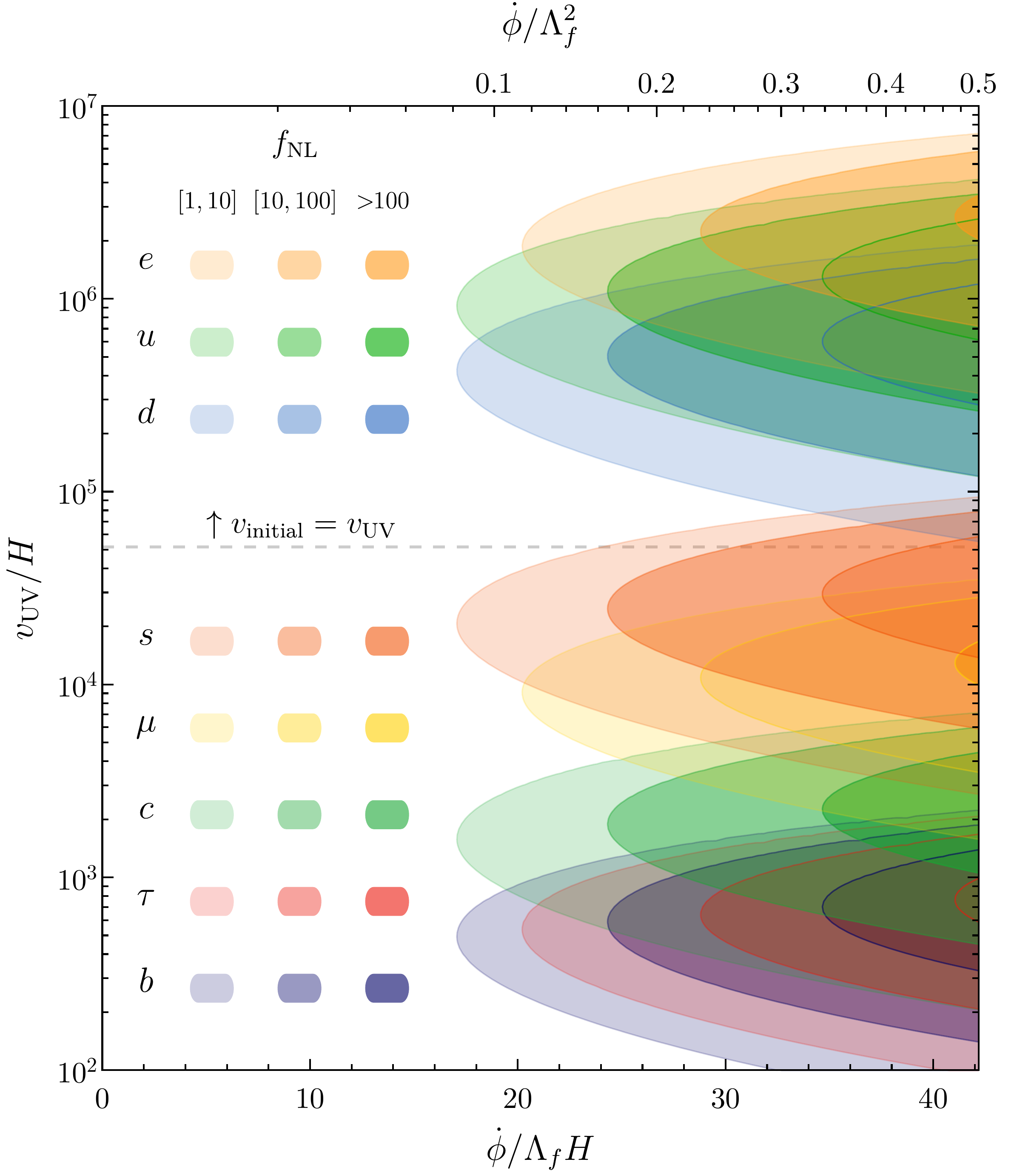}
\caption{The size of $\fNL$ that can generated by the production of SM fermions ($f \in \left\{b,\,\tau,\,c,\,\mu,\,s,\,d,\,u,\,e\right\}$ from bottom to top) during inflation as a function of the Higgs vev $\vUV$ in Hubble units and the strength of coupling $\lt=\dot\phi/\Lf H$ between the inflation and the SM fermions (we recall that we are setting $c_{f_i}=1$). 
In the range of the parameter space where $H \ll \left|\lUV\right|^{1/2}\vUV \simeq 0.1 \vUV$ the fluctuations of the Higgs field around $\vUV$ are negligible, and a large $\fNL \lesssim 100$ can be generated by SM fermions when $m_f/H$ is $\mathcal{O}(1)$. 
In the region above the dashed gray line, the observable signature requires the Higgs field to be in the UV minimum $\vUV$ at the beginning of the observable $\mathcal{O}(60)$ last $e$-folds of inflation. 
In large portions of the parameter space, there is the possibility of observing more than one fermion contributing with $\fNL\gtrsim 10$. We entertain the possibility of $\fNL \approx 100$ to account for the potential suppression coming from the different shape of non-Gaussianity compared to commonly studied templates.
In particular, as a result of the infamous $b-\tau$ unification in the SM, there is the possibility of observing both with strength $\fNL\gtrsim 100$. 
The couplings between the SM fermions and the inflaton $c_i$ are chosen to be the same for all SM species (see Appendix~\ref{app:inflaton} for a detailed discussion). 
The Yukawa couplings of the fermions are evaluated at a scale of $10^{13}$ GeV. 
}
\label{fig:signalf}
\end{figure}

In Sec.~\ref{sec: ccsignature}, we discussed the non-Gaussian signature that can arise from a single SM particle with shift-symmetric couplings with the inflaton. 
As discussed in Sec.~\ref{sec:potentialinflation}, the SM fermions provide a natural comb to scan the Hubble scale during inflation. 
In Fig.~\ref{fig:signalf}, we show the particles ($b,\,\tau,\,c,\,\mu,\,s,\,d,\,u,\,e$) that can contribute significantly to a non-Gaussian bispectrum of the inflationary perturbations. 

If the Hubble scale during inflation lies in the range ($10^{11}\GeV \approx v_{\lambda=0} \lesssim H  \ll \vUV \lesssim 10^{16} \GeV$), independently from the exact value of $H$, there is at least one SM particle which can induce an $\fNL \gtrsim 10$. 
For much smaller Hubble scales ($H \lesssim v_{\lambda=0}$), the first generation of SM fermions ($d,\, u,\, e$) can also contribute to an observable signal. 
In this case, the existence of the UV minimum $\vUV$ is not enough: if the Higgs field starts near the origin, then it cannot go beyond the barrier during the observable $\mathcal{O}(60)$ $e$-folds of inflation. If instead the Higgs field lies in the UV minimum $\vUV$ at the beginning of the last $60$ $e$-folds of inflation, we still get an observable effect%
\footnote{In the case where $H\ll v_{\lambda=0}$ and the electroweak symmetry is unbroken at the start of the observable $\mathcal{O}(60)$ $e$-folds of inflation, surprisingly, a unique signature can also arise. 
We study this case in a companion paper~\cite{Hook:2019}.}. 
In a wide range of the parameter space, as a result of the close proximity of the Yukawa couplings of the SM fermions (see Fig.~\ref{fig: running SM}), one can potentially see the effect of more than one fermion. 
In particular, as a result of the infamous $b-\tau$ unification\footnote{We thank Prateek for very emotional discussions.} in the SM, both fermions can contribute with an $\fNL \gtrsim 100$ in some range of the parameter space.

The possibility of observing multiple fermions is very important for distinguishing our signal from that of a generic fermion that couples with the inflaton in the same way. 
From the observation of the amplitude and the frequency of the oscillatory signal, it is possible to extract two independent quantities: the mass of the fermion in Hubble units $\mt$ and the strength of coupling in Hubble units $\lt$. 
If the fermions were to come from the SM, the ratio of the measured $\mt$ should be equal to the ratio of the Yukawa couplings of the two fermions, as both the Hubble scale and the Higgs vev $\vUV$ cancel out:
\begin{equation}
\frac{\mt_i}{\mt_j} = \frac{y_i}{y_j} \,.
\end{equation}

Besides a simultaneous measurement of two fermions, one has the chance to observe fermions in combination with a non-zero scalar-to-tensor ratio $r$ in the case of high scale inflation. 
Future measurement of $r$ can potentially extend the sensitivity to the Hubble scale to as low as $8\times 10^{12} \GeV$ (assuming a sensitivity of $\sigma (r) \sim 10^{-3}$ with CMB-S4~\cite{Abazajian:2016yjj}). 
The fermions that can possibly be simultaneously measured through their non-analytical contribution to the bispectrum in case of an observable scalar-to-tensor ratio can only be the bottom quark and the $\tau$-lepton (see Fig.~\ref{fig:  plane v H})~\footnote{Simultaneously measuring bottom, $\tau$ and charm would be more interesting as it can provide insight on some of the harder to probe scenarios~\cite{ArkaniHamed:2004yi,Arvanitaki:2012ps}}. 
As a result of a good measurement of their mass ratio, we could know quite precisely the value of the scale $\vUV$, due to the close numerical vicinity of $y_b$ and $y_\tau$ at high energies.
This exciting possibility would have extremely interesting implications for Grand Unification Theories~\cite{Georgi:1974sy,Pati:1974yy,Dimopoulos:1981zb,Dimopoulos:1981yj}, String Theory and other UV dynamics.

\begin{figure}[h!] \centering
\includegraphics[width=0.8\textwidth]{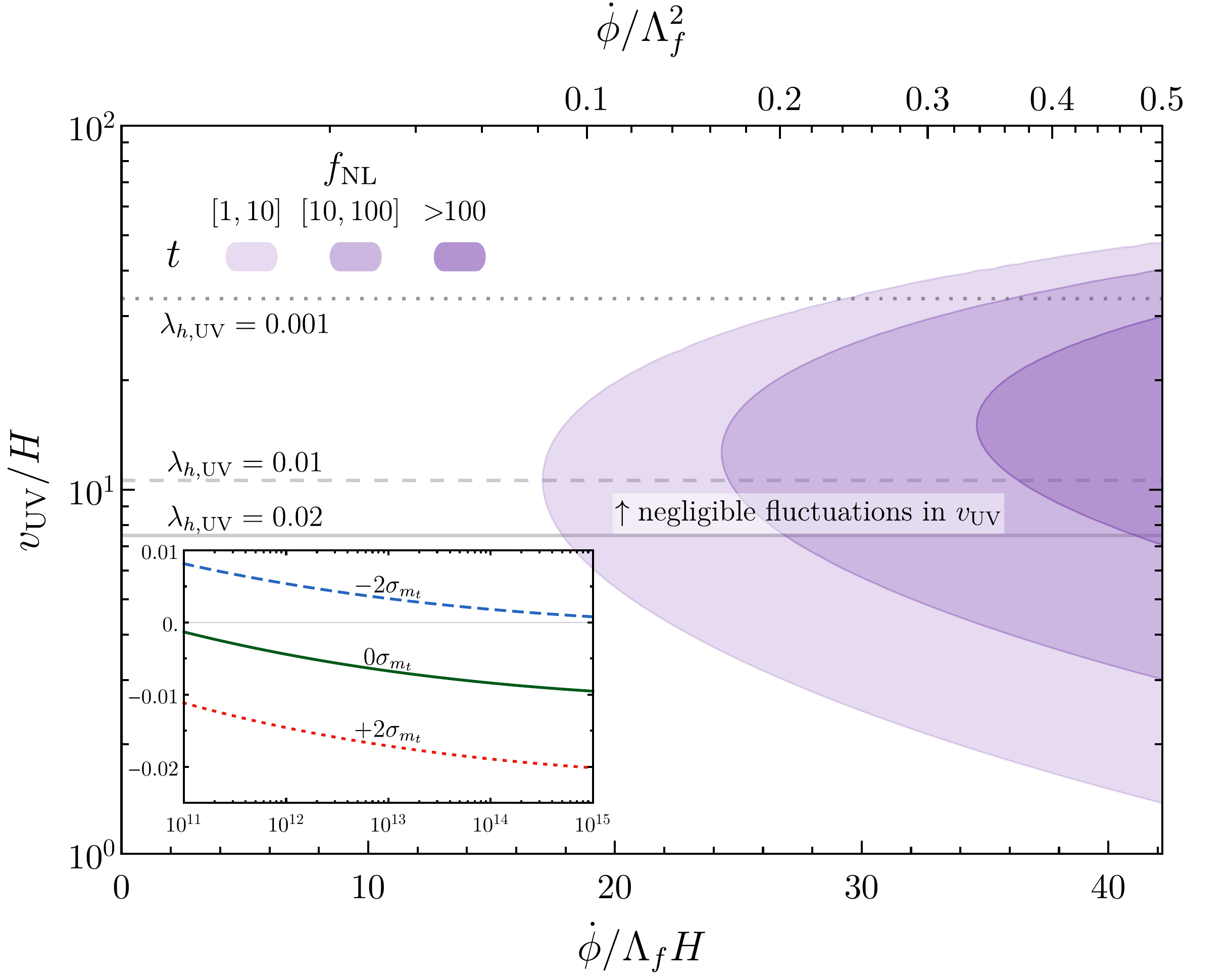}
\caption{The same plot as Fig.~\ref{fig:signalf}, specialized to the top quark. Above the dotted, dashed and solid gray lines, the Higgs fluctuations are exponentially suppressed for Higgs quartics $\lUV$ larger than the indicated value. 
The inset shows the running of the Higgs quartic $\lUV$ as a function of the RG energy scale, and its dependence on the uncertainty of the measurement of the top quark mass at the LHC (see \cite{Franciolini:2018ebs} for the reference values for $m_t$).). 
The green solid line corresponds the central value while the blue dashed and the red dotted lines show the $\pm 2\sigma$ contours.}
\label{fig:signaltop}
\end{figure}

A special case is the signature from the top quark, which generates the strongest signal when $\vUV/H \lesssim 100$ (see Fig.~\ref{fig:signaltop}). However, as it is apparent from Fig.~\ref{fig: plane v H}, most of the signal from the top quark (when $m_t/H \approx 1$) lies in the range where the Higgs field can quantum fluctuate during inflation. 
The region where this fluctuation can be important depends very strongly on the value of the Higgs mass and, as a result, the quartic coupling of the Higgs. 
When $\lUV^{1/2}\vUV/H \gtrsim 1$, the signal of the top quark is the same as the other SM fermions. 
On the other hand, when $\lUV^{1/2}\vUV/H \lesssim 1$, the Higgs field value and, consequently, the SM fermion masses, could have significant fluctuations during inflation. 
Therefore, different patches of the universe can potentially have signals with different amplitude. 
We leave a study of how to compute and extract this signal from data to future work. 

The signature that we studied in this paper has the rare features that the signal strength is largest when $k_3 \sim k_1/\tilde{\lambda}$, deep in the squeezed limit, and the oscillatory part of the signal has a frequency that is much larger than the mass of the populated particle in Hubble units.
Both properties emerge as a result of the very uncommon ``redshifting'' of the fermions during inflation in presence of the modified dispersion relation of Eq.~\eqref{eq: dispersion relation}, while when $k_3 \simeq k_1$, such enhancement disappears and the signal strength is $\fNL \sim \frac{\dot{\phi}}{\Lf^2}\frac{H}{\Lf}/(2\pi)^4 \ll 1$ from the UV contribution of the loop diagram in Fig.~\ref{fig:dtriangle}~\cite{Chen:2018xck}. 
These features imply that the search for these signatures will greatly benefit from measurements of the large scale structure of the Universe~\cite{Munoz:2015eqa,Alvarez:2014vva,Shandera:2010ei}, and in particular, the upcoming program of 21cm cosmology~\cite{Meerburg:2016zdz}. 
This will provide us with potentially more modes than the CMB, as well as a 3D map of the density perturbations in the Universe, which will be important to uncover small signals, and to precisely measure the oscillation frequency. 
We postpone a more detailed study of the observability of our proposed signature to future work.

We would like to close with a final question for the reader. Would we take a different view about the electroweak hierarchy problem if we were to find a new minimum in the Higgs potential? What if we found a wealth of them?

\acknowledgments{
We thank Soubhik Kumar and Zhong-Zhi Xianyu for very helpful and instructive discussions. 
We also thank Prateek Agrawal,
Asimina Arvanitaki, 
Masha Baryakhtar, 
Arushi Bodas,
Neal Dalal, 
Savas Dimopoulos, 
Victor Gorbenko, 
Matthew Johnson, 
Gustavo Marques-Tavares, 
Mehrdad Mirbabayi, 
Moritz M\"unchmeyer, 
Antonio Riotto, 
Leonardo Senatore, 
Raman Sundrum, 
Ziqi Yan 
for many useful discussions. 
JH and DR thank the Stanford Institute for Theoretical Physics, and AH and JH thank MIAPP and KITP Santa Barbara, for generous hospitality during the completion of this work.
This research was supported in part by Perimeter Institute for Theoretical Physics. Research at Perimeter Institute is supported in part by the Government of Canada through the Department of Innovation, Science and Economic Development Canada and by the Province of Ontario through the Ministry of Economic Development, Job Creation and Trade.  
AH is supported in part by the NSF under Grant No. PHY-1620074 and by the Maryland Center for Fundamental Physics (MCFP). This research was supported in part by the National Science Foundation under Grant No. NSF PHY-1748958.
}

\appendix

\section{Calculation of the squeezed non-Gaussianity}
\label{app: App_NG}
In this Appendix we show in some detail our estimate of the non-Gaussianity in the squeezed limit.
The outline of the calculation and the notation closely follow Ref.~\cite{Chen:2018xck}, with some differences in the late time expansion of the fermion wavefunctions and in the final result for $\fNL$.

\subsection{In-In formalism}
The calculation of correlation functions in cosmology requires a different treatment with respect to the familiar one of quantum field theory.
The key differences are that we usually want to compute correlation functions of fields evaluated at a fixed time, and not at asymptotically large times. 
Also the Hamiltonian describing the field fluctuations depends on time, because of the time dependence of the background fields. 
Finally, the condition on the fields are imposed at very early times when (in the inflationary context) the relevant modes are well within the Hubble radius and we recover the standard solutions in Minkowski space.

We refer to \cite{Weinberg:2005vy,Chen:2010xka,Chen:2017ryl} for a detailed treatment of the in-in formalism and references to the original literature.
We collect here just some relevant formul\ae\ to set the stage for the remainder of the calculation.

The expectation value for an operator $Q(\tau)$ built out of the fields of the model evaluated at a time $\tau$ can be computed as \cite{Weinberg:2005vy,Chen:2010xka}
\begin{equation}
\label{eq: in-in hamiltonian}
\left\langle Q(\tau) \right\rangle = \Big\langle \Omega \Big| 
\left[ \overline T  \exp\left(i \int_{\tau_0}^\tau H_I(\tau') \mathrm d \tau'\right) \right]
Q^I(\tau)
\left[ T  \exp\left( -i \int_{\tau_0}^\tau H_I(\tau'') \mathrm d \tau''\right) \right]
\Big|\Omega \Big\rangle
\end{equation}
where $Q^I$ and $H_I$ are the operator $Q$ and the Hamiltonian in the interaction picture, $|\Omega\rangle$ is the vacuum state at an early time $\tau_0$, and $T$, $\overline T$ denote the time- and anti-time-ordering operators.

The expectation value can be equivalently formulated in terms of a path integral. 
If we denote the fields $\varphi$ of the Lagrangian $\mathscr L$ with a subscript $\oplus$ and $\ominus$ (standing respectively for $+1$ and $-1$, and also denoted generically by a so called in-in index $\oa_i$) depending on whether the fields should be time- or anti-time-ordered (that is, depending on which of the two time evolution operators in Eq.~\eqref{eq: in-in hamiltonian} the fields come from), one can rewrite the expectation value through functional derivatives of a generating functional \cite{Chen:2017ryl}:
\begin{gather}
Z[J_\oplus, J_\ominus] = \int \mathscr D \varphi_\oplus \mathscr D \varphi_\ominus \, \exp\left[i\int_{\tau_0}^{\tau_f} \mathrm d \tau' \mathrm d^3x \left(\mathscr L[\varphi_\oplus] -\mathscr L[\varphi_\ominus] \right) +J_\oplus \varphi_\oplus - J_\ominus \varphi_\ominus \right]  \\
\langle \varphi_{\oa_1}(\tau,\vec x_1) \cdots \varphi_{\oa_n}(\tau,\vec x_n) \rangle = 
\frac{\delta}{i\oa_1 \delta J_{\oa_1}(\tau, \vec x_1)}\cdots
\frac{\delta}{i\oa_n \delta J_{\oa_n}(\tau, \vec x_n)}
Z[J_\oplus,J_\ominus]\Big|_{J_\oplus=J_\ominus=0}.
\end{gather}
Within the usual perturbative treatment, we expand the exponential of the action and we keep the leading order terms. 
Each occurrence of the action leads to a vertex carrying a time integral, which will have to be eventually performed in the calculation of the expectation value.
Any vertex is characterized by a given in-in index $\oa$, and the final answer requires us to sum over $\oa=+1,\,-1$.
We refer the reader to \cite{Chen:2017ryl} for a more comprehensive exposition of Schwinger-Keldish diagrammatic calculations.

\subsection{Fermion loop amplitude}
\begin{figure}[h!] \centering
\includegraphics[width=1\textwidth]{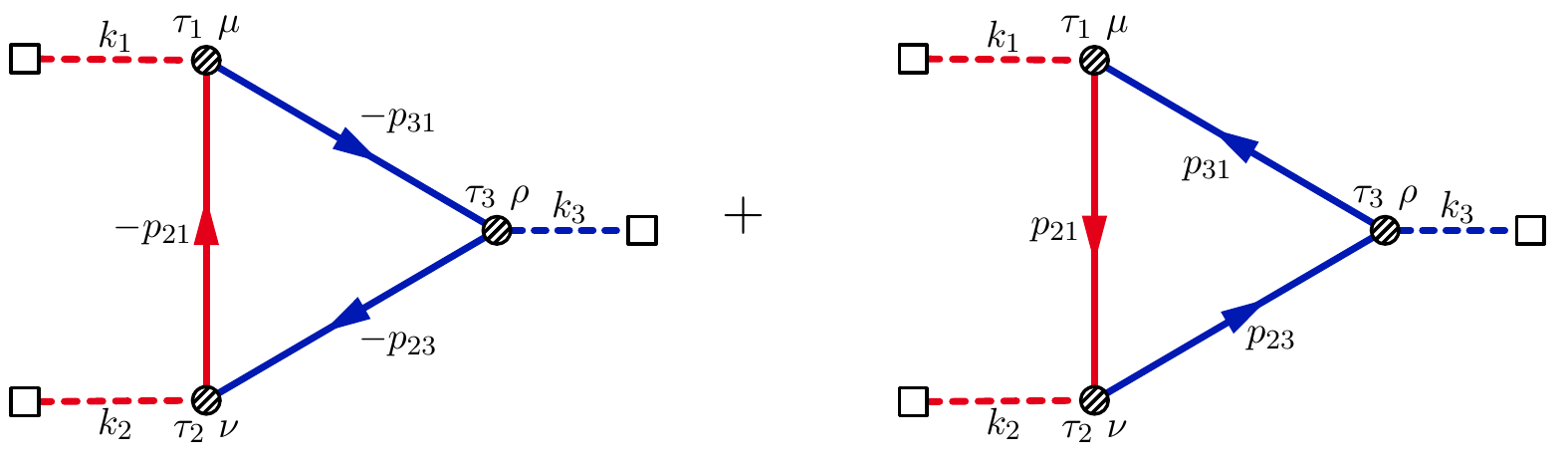}
\caption{Feynman diagrams showing the main contributions to the 3-point function for the inflaton. 
The inflaton and fermion lines highlighted in red (blue) highlight the hard (resp.\ soft) momenta in the squeezed limit.}
\label{fig:dtriangle}
\end{figure}

The main contribution to 3-point function of the inflaton comes from loop diagrams with the exchange of a SM fermion.
The two contributing diagrams are shown in Fig.~\ref{fig:dtriangle}. 
\newline
We denote with a white square the inflaton field evaluate at late times (on the $\tau=0$ hypersurface in Fig.~\ref{fig:dinf}), and with a hatched circle the vertices, to understand the sum the two contributions for each value of the in-in index associated to the vertex. 
We denote in-in indices by $\oa,\ob=\pm 1$ and $\oplus, \ominus$, in order to distinguish them from the fermion helicity indices $a,b = \pm 1$.

\paragraph{Vertices}
The relevant terms involving the fermions fields in the Lagrangian density are (in four-component notation)
\begin{equation}
\mathscr L = \sqrt{-g} \left[ \overline{\tf_i} i\slashed{D} \tf_i -m \,\overline{\tf_i} \tf_i - \frac{c_{f_i}}{\Lf} \overline{\tf_i} (\slashed D \phi)\gamma_5 \tf_i \right] \,,
\end{equation}
where $\slashed D = \partial_\mu e^\mu_a \gamma^a$ (we understand the gauge covariant derivative) and $e^\mu_a$ is the vierbein connection. Specialising to de Sitter metric (which corresponds to the background metric during inflation up to corrections suppressed by the slow-roll parameters), we have $\sqrt{-g} = a^4$, $\slashed D = a^{-1}\partial_\mu \delta^\mu_a \gamma^a$.
After performing the redefinition $f_i=a^{3/2}\tf_i$ to factor out the dilution of the fermion wavefunction due to the spacetime expansion, we get
\begin{equation}
\mathscr L = \overline{f_i} \, i\gamma^\mu \partial_\mu f_i - (a\,m)\, \overline{f_i}f_i -\frac{c_{f_i}}{\Lf} \partial_\mu \phi \,\overline{f_i} \gamma^\mu \gamma_5 f_i \,.
\label{eq: Lagr f}
\end{equation}
The interaction term, when evaluated on the inflaton background, gives 
$\tfrac 1 \Lf \partial_\mu\phi = 
 \tfrac 1 \Lf \partial_\tau\phi \, \delta_{\mu 0} = 
 a \lambda \, \delta_{\mu 0}$
where $\lambda = \dot\phi/\Lf$.

The interaction vertex appearing in the two diagrams in Fig.~\ref{fig:dtriangle} is
\begin{equation}
- \frac{c_{f_i}}{\Lf}\, \partial_{\mu}{\phi} \, \overline{f_i} \gamma^{\mu} \gamma^5 f_i
\end{equation}
in four-component spinor notation, or $c_{f_i} /\Lf\, \partial_{\mu}{\phi} \, (
f_{L,i}^\dagger \overline \sigma^{\mu} f_{L,i} +
f_{R,i}^\dagger \overline \sigma^{\mu} f_{R,i})$ in two-component spinor notation where $f_i = \begin{pmatrix} f_{L,i} \\ f_{R,i}^\dagger \end{pmatrix}$.
In the remainder of this Appendix we work in the two-component notation. 

For the calculation of these Feynman diagrams we incorporate the derivative into the inflation field. 
Each vertex carries then a factor
\begin{equation}
(\oa i) \int_{-\infty}^0 \mathrm d \tau\, \frac{c_{f_i}}{\Lf}\, \overline \sigma ^{\mu \dot \alpha \alpha}
\label{eq: vertex}
\end{equation} 
where $\oa=\pm 1$ is the in-in index related to the whether the vertex comes from the time or anti-time ordered product. 
From now on, we set $c_{f_i}=1$; it can be easily restored by replacing each occurrence of $1/\Lf$ by $c_{f_i}/\Lf$.

\paragraph{External inflaton lines}
The correlator of the inflaton field $\partial_\mu\phi$ from each vertex with the same field evaluated at late times $\tau=0$ is computed by taking the derivative of the so called boundary-to-bulk correlator $G_{\oa}(\vec k,\tau)=\langle \phi^{(*)}(\tau,\vec k) \phi(0,\vec k) \rangle$, where the in-in index  $\oa$ (with the evaluation of $\phi$ for $\oa=+1$ and $\phi^*$ for $\oa=-1$) distinguishes whether the vertex comes from the time or anti-time ordered product:
\begin{equation}
G_{\oa}(\vec k,\tau) = \frac{H^2}{2k^3}(1- i \oa k \tau) e^{i\oa k \tau} \,,
\end{equation}
where $\oa= \pm1$. Then for each external leg with Lorentz index $\mu$ and in-in index $\oa$ we get the following function:
\begin{equation}
F_{\mu\oa}(\vec k,\tau) = 
\begin{pmatrix}
\partial_\tau G_\oa \\ i\vec k G_\oa
\end{pmatrix} = 
\frac{H^2}{2k^3}
\begin{pmatrix}
k^2 \tau \\ i\vec k (1-i \oa k \tau)
\end{pmatrix}
e^{i \oa k \tau}
\label{eq:Ffunction}
\end{equation}

\paragraph{Fermion loop}
Let us fix the notation for the fermion wavefunction, postponing a more detailed discussion about the solution to Sec.~\ref{app: fermion mode functions}.
We switch to the two-component notation, denoting both $f_{L,i}$ and $f_{R,i}$ by $\psi$, and we expand $\psi$ into eigenfunctions of the 3-momentum,
\begin{equation}
\psi_{\alpha} (\tau, \vx) = \int \frac{\mathrm d^3 k}{(2\pi)^3} \sum_{s=\pm 1} \left[ 
\xi_{\alpha,s} (\tau,\vk) a_s(\vk) e^{i\vk\cdot \vx} + 
\chi_{\alpha,s} (\tau,\vk) a_s^\dagger(\vk) e^{-i\vk\cdot \vx}
\right] \,,
\label{eq: def f}
\end{equation}
where $\alpha$ is a spinor index and $s$ the helicity index, $a_s$ and $a_s^\dagger$ are the annihilation and creation operators satisfying $[a_s(\vk),a^\dagger_{s'}(\vk')]=(2\pi)^3 \delta_{s\, s'}\, \delta^{(3)}(\vk-\vk')$.
We denote the positive and negative frequency components of the fermion $\psi$ by $\xi$ and $\chi$, with respective mode functions $u_s$ and $v_s$ (we define $v$ through $\chi^\dagger$ so that both $u$ and $v$ have positive energy) and helicity eigenstate spinors $h_{\alpha, s}$:
\begin{equation}
\begin{aligned}
\xi_{\alpha,s}(\tau, \vec k) & = \sum_{s=\pm 1} u_s(k\tau)  \, h_{\alpha, s} (\vec k) \,, \\
\chi_s^{\dagger\,\dot \alpha}(\tau, \vec k) & = \sum_{s=\pm 1} v_s(k\tau) \, h_s^{\dagger\, \dot \alpha} (\vec k)  \,.
\end{aligned}
\label{eq: def xi chi}
\end{equation}
The normalisation condition for the helicity eigenfunctions are
\begin{equation}
\vec \sigma\cdot \vec k \, h_s(\vec k) = s k \, h_s(\vec k) \,, \quad
h_s^\dagger (\vec k) h_{s'} (\vec k) = \delta_{s\, s'} \,, \quad
\sum_{s=\pm 1} h_s(\vec k) h_s^\dagger(\vec k) =1 \,,
\end{equation}
and are satisfied by the following expressions for $h_s(\vec k)$ where $\vec k = k(\sin \theta \cos \phi, \sin \theta \sin \phi, \cos \theta)$:
\begin{equation}
h_+(\vec k) = \begin{pmatrix} \cos \tfrac{\theta}{2} \\ e^{i\phi}\sin \tfrac \theta 2 \end{pmatrix} \,, \quad
h_+(\vec k) = \begin{pmatrix} -e^{-i\phi}\sin \tfrac{\theta}{2} \\ \cos \tfrac \theta 2 \end{pmatrix} \,.
\end{equation}
We postpone to Sec.~\ref{app: fermion mode functions} a derivation of the fermion mode functions $u_\pm$, $v_\pm$ and dispersion relations.
The propagators appearing in the amplitude associated to the diagrams in Fig.~\ref{fig:dtriangle} are of the type $\langle f_{\alpha}(\tau_1,\vec k) f^{\dagger\dot \beta}(\tau_2,\vec k)\rangle$. 
We denote them by $D_{\oa \ob\alpha}^{\ \ \ \ \dot \beta}(\vec k, \tau_1,\tau_2)$ and take the following form depending on the in-in indices $\oa,\ob$ associated to the two fermion functions. 
If the two fields come both from the time or anti-time ordered product, then an Heaviside function enforces the ordering.
\begin{equation}
\begin{aligned}
D_{\oplus \oplus \alpha}^{\ \ \ \ \ \dot \beta}(\vec k,\tau_1,\tau_2) & = \xi_\alpha(\tau_1,\vec k) \xi^{\dagger\dot\beta}(\tau_2,\vec k) \theta(\tau_1-\tau_2) - \chi^{\dagger\dot\beta}(\tau_2,\vec k) \chi_{\alpha}(\tau_1,\vec k) \theta(\tau_2-\tau_1) \\
D_{\oplus \ominus \alpha}^{\ \ \ \ \ \dot \beta}(\vec k,\tau_1,\tau_2) & = - \chi^{\dagger\dot\beta}(\tau_2,\vec k)   \chi_\alpha(\tau_1,\vec k) \\
D_{\ominus \oplus \alpha}^{\ \ \ \ \ \dot \beta}(\vec k,\tau_1,\tau_2) & = \xi_\alpha(\tau_1,\vec k) \xi^{\dagger\dot\beta}(\tau_2,\vec k) \\
D_{\ominus \ominus \alpha}^{\ \ \ \ \ \dot \beta}(\vec k,\tau_1,\tau_2) & =  - \chi^{\dagger\dot\beta}(\tau_2,\vec k) \chi_{\alpha}(\tau_1,\vec k) \theta(\tau_1-\tau_2) +\xi_\alpha(\tau_1,\vec k) \xi^{\dagger\dot\beta}(\tau_2,\vec k) \theta(\tau_2-\tau_1) 
\end{aligned}
\end{equation}
The two diagrams shown in Fig.~\ref{fig:dtriangle} give the two following fermion traces (we include here the Pauli matrices coming from each vertex in Eq.~\eqref{eq: vertex})
\begin{equation}
\begin{aligned}
\mathscr T^{\mu\nu\rho}_{\oa\ob\oc} =& -\text{tr}\left[
\overline \sigma^{\mu\dot\alpha \alpha} D_{\oa \ob \alpha \dot \beta}(\vec p_{12},\tau_1,\tau_2)
\overline \sigma^{\nu\dot\beta \beta} D_{\ob \oc \beta \dot \gamma}(\vec p_{23},\tau_2,\tau_3)
\overline \sigma^{\rho\dot\gamma \gamma} D_{\oc \oa \gamma \dot \alpha}(\vec p_{31},\tau_3,\tau_1)
\right] \\
&  -\text{tr}\left[
\overline \sigma^{\mu\dot\alpha \alpha} D_{\oa \oc \alpha \dot \gamma}(-\vec p_{31},\tau_1,\tau_3)
\overline \sigma^{\rho\dot\gamma \gamma} D_{\oc \ob \gamma \dot \beta}(-\vec p_{23},\tau_3,\tau_2)
\overline \sigma^{\nu\dot\beta \beta} D_{\ob \oa \beta \dot \alpha}(-\vec p_{12},\tau_2,\tau_1)
\right]
\end{aligned}
\label{eq:trace T}
\end{equation}

\paragraph{Final amplitude}
We can finally write down the full expression for the fermion loop contribution to the three-point function:
\begin{multline}
\left\langle\delta \phi (\vec k_1) \delta \phi ( \vec k_2) \delta \phi (\vec k_3)\right\rangle = \\
\sum_{\oa,\ob,\oc =\pm 1} \oa\ob\oc \left(\frac{i}{\Lambda}\right)^3 \iiint_{-\infty}^0 \mathrm{d} \tau_1 \mathrm{d} \tau_2 \mathrm{d} \tau_3 F_{\mu\oa}(\vec k_1,\tau_1) F_{\nu\ob}(\vec k_2,\tau_2) F_{\rho\oc}(\vec k_3,\tau_3)
\int \frac{\mathrm d^3q}{(2\pi)^3} \mathscr T^{\mu\nu\rho}_{\oa\ob\oc},
\label{eq:full amplitude}
\end{multline}
where the external lines $F_{\mu \oa}$ and the fermion trace $\mathscr T^{\mu\nu\rho}_{\oa\ob\oc}$ are defined in Eq.~\eqref{eq:Ffunction} and \eqref{eq:trace T}, and $\vec q \equiv \vec p_{12}$.\\
A full analytical solution to the time and momentum integrals in Eq.~\eqref{eq:full amplitude} is not possible, due to complicated form of the fermion wavefunctions.
In the next sections we discuss the relevant approximations that allow us to obtain an estimate of this contribution in the squeezed limit $k_1,k_2 \gg k_3$ and $\lambda \gg m$.

\subsection{Fermion mode functions and dispersion relations}
\label{app: fermion mode functions}
We derive now the equations of motion and their solution for the fermions.
Starting from the Lagrangian density in Eq.~\eqref{eq: Lagr f} evaluated on the inflaton background%
\footnote{For this section, and in particular for the derivation of the fermion mode functions, we write the inflaton coupling to the fermions as $a \dot\phi /\Lf = a \lambda$ to highlight the time dependence, whereas for the rest of the computation we write it in terms of the conformal time $\partial_\tau\phi$ to directly compute the external inflaton lines $F_{\mu\oa}$.}, and defining the mode functions as in Eqs.~\eqref{eq: def f} and \eqref{eq: def xi chi}, we obtain the following equations of motion for $u_s$ and $v_s$:
\begin{equation}
\begin{aligned}
i u_\pm' +(\pm k - a\lambda) u_\pm & = a m v_\pm \\
i v_\pm' -(\pm k - a\lambda) v_\pm & = a m u_\pm \\
\end{aligned} 
\end{equation}
These equations can be rewritten into two separate second order differential equations for $u$ and $v$:
\begin{equation}
\begin{aligned}
u_\pm'' - aH u_\pm' +\left[(k \mp a\lambda)^2 +a^2 m^2 \pm iaHk\right] u_\pm & = 0 \\
v_\pm'' - aH v_\pm' +\left[(k \mp a\lambda)^2 +a^2 m^2 \mp iaHk\right] v_\pm & = 0 
\end{aligned} 
\end{equation}
These equations of motion show explicitly the dispersion relation introduced and discussed in Eq.~\eqref{eq: dispersion relation} in Sec.~\ref{sec: fNL calculation} (obtained in the approximation $\lambda,m \gg H$).
Their solutions are given by the Whittaker functions $W$:
\begin{equation}
\label{eq: fermion Whittaker}
\begin{aligned}
u_+ (k\tau) & = \frac{\mt \, e^{\pi \lt/2}}{\sqrt{-2 k \tau}} W_{-\frac 12 - i \lt, i \mut} (2 i k \tau)\,, &
v_+ (k\tau) & = \frac{i \, e^{\pi \lt/2}}{\sqrt{-2 k \tau}} W_{+\frac 12 - i \lt, i \mut} (2 i k \tau)\,, \\
u_- (k\tau) & = \frac{i \, e^{-\pi \lt/2}}{\sqrt{-2 k \tau}} W_{+\frac 12 + i \lt, i \mut} (2 i k \tau)\,, &
v_- (k\tau) & = \frac{\mt \, e^{-\pi \lt/2}}{\sqrt{-2 k \tau}} W_{-\frac 12 + i \lt, i \mut} (2 i k \tau)\,.
\end{aligned}
\end{equation}
We collect here some useful formul\ae\ to treat the late time expansion of the Whittaker functions (see e.~g.~\cite{dlmf}).
There is a connection formula between the Whittaker functions $W$ and $M$,
\begin{equation}
W_{\kappa,\mu} (z) = \frac{\Gamma(2\mu)}{\Gamma(\frac 12 +\mu-\kappa)} M_{\kappa,-\mu}(z) + (\mu\leftrightarrow -\mu) \,.
\end{equation}
A useful formula to expand the Whittaker functions $M$ around $z=0$ for $-2\mu\notin \mathbb N$ (which is always the case for us) is
\begin{equation}
M_{\kappa, \mu} (z) = e^{-\frac 12 z} z^{\frac 12 +\mu}  \sum_{n=0}^\infty \frac{(\frac 12 +\mu-\kappa)_s}{(1+2\mu)_s \, s!} \, z^s 
= e^{-\frac 12 z} z^{\frac 12 +\mu}  \left( 1+\frac{\frac 12 +\mu-\kappa}{1+2\mu} z +\dots \right)  \,,
\end{equation}
where $(a)_s=\Gamma(a+s)/\Gamma(a)$ is the Pochhammer symbol. 
We have written the first subleading term in the expansion around $z=0$ because we want to check the goodness of the late time expansion up to $-k\tau\lesssim \mut$, which is the relevant range for one of the time integrals in our calculation.
Some properties of the Gamma functions will be useful:
\begin{equation}
\Gamma(1+ia)=ia\, \Gamma(ia) \,, \qquad |\Gamma(\pm i \mu)| \overset{\mu\to\infty}{\to}  \frac{1}{\sqrt{|\mu|}} e^{-\pi |\mu|/2 } \,.
\end{equation}
From these equations, we can derive the late time expansion of the fermion mode functions, in the limit of small $k \tau$.
We keep the first subleading term in the expansion, and we underline the terms for which the leading term is \textit{not} a good approximation for $-k\tau \lesssim \mut$ in the limit $\lt \gg \mt$, but just until $-k\tau \lesssim 1$.
We also highlight the terms that are exponentially suppressed in the limit $\lt\gg\mt$ which is relevant for us.
\begin{equation}
\label{eq: fermion Whittaker late}
\begin{aligned}
u_+ (k\tau) & = \mt \, e^{-i k \tau} e^{\pi \lt/2} e^{-\frac{i\pi}{4}}
  \left[ \frac{e^{-\pi \mut/2} \Gamma(2i\mut)}{\Gamma(1+i(\mut+\lt))} (-2k\tau)^{-i \mut} 
  \left( 1+ \frac{1-i(\mut-\lt)}{1-2i\mut} 2ik\tau\right)  
  + \underline{(\mut \leftrightarrow -\mut)}\right] \\
u_- (k\tau) & = e^{-i k \tau} e^{-\pi \lt/2} e^{\frac{i\pi}{4}}
  \Bigg[ \underset{\overset{\lt\gg\mt}{\to} e^{-3\pi\lt/2}}{\underbrace{\frac{ e^{-\pi \mut/2}\Gamma(2i\mut)}{\Gamma(i(\mut-\lt))}}} (-2k\tau)^{-i \mut} 
  \underline{\left( 1+ \frac{-i(\mut+\lt)}{1-2i\mut} 2ik\tau\right)}
  + (\mut \leftrightarrow -\mut)\Bigg] \\
v_+ (k\tau) & = e^{-i k \tau} e^{\pi \lt/2} e^{\frac{i\pi}{4}}
  \left[\frac{ e^{-\pi \mut/2} \Gamma(2i\mut)}{\Gamma(i(\mut+\lt))} (-2k\tau)^{-i \mut} 
  \left( 1+ \frac{-i(\mut+\lt)}{1-2i\mut} 2ik\tau\right)  
  + \underline{(\mut \leftrightarrow -\mut)} \right] \\
v_- (k\tau) & = \mt \, e^{-i k \tau} e^{-\pi \lt/2} e^{-\frac{i\pi}{4}}
  \Bigg[ \underset{\overset{\lt\gg\mt}{\to} e^{-3\pi\lt/2}}{\underbrace{\frac{ e^{-\pi \mut/2} \Gamma(2i\mut)}{\Gamma(1+i(\mut-\lt))}}} (-2k\tau)^{-i \mut} 
  \underline{\left( 1+ \frac{1-i(\mut+\lt)}{1-2i\mut} 2ik\tau\right)}  
  + (\mut \leftrightarrow -\mut)\Bigg]
\end{aligned}
\end{equation}
We'll comment about the underlined terms in Sec.~\ref{sec: time integrals}, when discussing how in our limit $\lt\gg\mt$ the time integral over $\tau_3$ selects only terms for which the early time expansion is valid up to $-k\tau \lesssim \mut$.

\subsection{Approximation for the momentum loop}
A full solution to the momentum and time integrals in the amplitude of Eq.~\eqref{eq:full amplitude} is very hard. We now motivate an approximation for the loop momentum integrals that allow us to perform the time integrals in the next section.

Looking at Fig.~\ref{fig:dinf} and \ref{fig:dtriangle}, we can see that the vertex at the early time $\tau_3$ involves a soft inflaton leg with momentum $k_3$, and two fermions which give the largest contribution to the signal when they have a momentum $p_{23}\tau_3 \sim p_{31}\tau_3 \sim\lt$ yielding a soft frequency of order $m$.
The integration over $\tau_3$ is dominated by $k_3 \tau_3 \sim \lt$ (as we will see in next section), so that the larger contribution to the 3-point function comes from configurations where $p_{23} \sim p_{31} \sim k_3$. 
We draw them accordingly with the same blue color in the Feynman diagrams.

The two produced fermions are then redshifted, and due to the dispersion relation \eqref{eq: dispersion relation} they quickly increase their energy to $\omega\sim \mu$. 
Thus the vertices at late times involve a hard momentum exchange (the inflaton legs have a standard dispersion relation), and the momentum flowing along the third fermion line is of order $p_{12}\sim k_1 \sim k_2$ (shown in red in the Feynman diagrams). 
We approximate therefore the corresponding propagator with the one in flat space.

We can finally write down an explicit parametrization for the internal momenta. 
We choose the following orientation for the momenta $\vec k_i$ (the orientation of $\vec k_3$ is not important for the result):
\begin{equation}
\vec k_1 = (0,0,k_1) \,, \qquad
\vec k_2 \sim -\vec k_1 = (0,0,-k_1) \,, \qquad
\vec k_3 = (0,0,k_3)  \,.
\end{equation}
The internal momenta satisfy the conditions $\vec q \equiv \vec p_{12}=\vec k_1 + \vec p_{23}$ and $\vec p_{31}-\vec p_{23} = \vec k_3$, and the most relevant regime for the fermion production is $|\vec p_{23}| \sim |\vec p_{31}| \sim |\vec k_3|$, so that the three vectors $\vec p_{31},\,\vec p_{23},\,\vec k_3$ approximately form an equilateral triangle and $\vec p_{12}\sim k_1$:
\begin{equation}
\vec p_{12} \simeq (0,0,k_1) \,, \quad
\vec p_{23} \simeq k_3 \left(\tfrac{\sqrt 3}{2} \cos\phi,\tfrac{\sqrt 3}{2} \sin\phi, -\tfrac 12\right) \,, \quad
\vec p_{31} \simeq k_3 \left(\tfrac{\sqrt 3}{2} \cos\phi,\tfrac{\sqrt 3}{2} \sin\phi, \tfrac 12\right) \,.
\end{equation}
This configuration is roughly obtained when $\vec p_{31}$ spans an annulus of radius and height of order $k_3$, so that we approximate the momentum integral to 
\begin{equation}
\int \frac{\mathrm d^3 q}{(2\pi)^3} \simeq \frac{k_3^3}{(2\pi)^3} \int_0^{2\pi} \mathrm d\phi \,.
\end{equation}

\subsection{Time integrals}
\label{sec: time integrals}
We have now all the ingredients to perform the calculation of the amplitude \eqref{eq:full amplitude}. 
The first step is computing the trace and the integral over $\phi$, leaving us with the time integrals over $\tau_3,\,\tau_2,\,\tau_1$ (where $\tau_3<\tau_2,\,\tau_1$).

The integrals are dominated by times of order $-k_i \tau_i \sim \mut$, and this can be shown as follows.
Let us denote $x_i\equiv k_i\tau_i$ for $i=1,2,3$. 
Each time integral includes an exponential $e^{\pm i x_i}$ from the external lines in Eq.~\eqref{eq:Ffunction}, together with a possible factor of $x_i$.
At early times, the Whittaker functions in the fermion mode functions go to zero and the integral does not receive a sizable contribution. 
We can then perform a late time expansion as in Eq.~\eqref{eq: fermion Whittaker late}, and we get a term $(-x_i)^{\pm i\mut}$ (and an exponential $e^{-ix_i}$ that we leave aside for a moment, since it does not affect this argument).
The oscillating integral
\begin{equation}
\int_{-\infty}^0 e^{\pm ix_i} (-x_i)^{\pm i \mut} \mathrm d x_i 
\end{equation}
can be related to $\Gamma$-functions by a contour integral. When the two signs are opposite, the integral is exponentially suppressed by $e^{-\pi \mu}$ compared to when the signs are the same.
We can also see from here how the oscillating feature $(k_3/k_1)^{\pm 2i\mut}$ emerges from the calculation. 
The fermion propagators for the soft lines involve fermion mode functions like $u_{\pm}(k_3 \tau_{1,2})$ which in the late time limit contain factors $(k_3\tau_{1,2})^{\pm i\mut}$. 
When writing the integral in dimensionless variables $x_i=k_i\tau_i$, a term $(k_3/k_1)^{\pm i\mut}$ is left. 
We obtain one such factor for both the integrals over $\tau_1$  and $\tau_2$.
The physical interpretation of this factor as related to the propagation of the two fermions was illustrated in Sec.~\ref{sec: fNL estimate}.

Returning back to the exponential $e^{-ik\tau}$ in the fermion functions \eqref{eq: fermion Whittaker late}, this is negligible for the mode functions involving late times, $u_\pm(k_3\tau_{1,2}),\,v_\pm(k_3\tau_{1,2})$, because $k_3\tau_1 =\tfrac{k_3}{k_1} (k_1\tau_1)\sim \tfrac{k_3}{k_1} \mut \ll \mut$.
It plays a role though in the integral over $\tau_3$ for the mode functions $u_\pm(k_3\tau_3)$, $v_\pm(k_3\tau_3)$, in order to have an expansion reliable up to $-k_3\tau_3 \lesssim \mut$. 
This aspect was not considered in \cite{Chen:2018xck}, and constitutes the difference between our results (whereas we have closely followed their procedure in the rest of the calculation).
When considering the integral over $\tau_3$, the largest contributions come from the pieces containing either%
\footnote{The enhancement of a particular helicity (in this case $s=+1$) is related to the sign of the inflaton coupling to fermions $\lambda =\dot\phi/\Lf$, which acts as a chemical coupling favouring the production of a given helicity mode.}
 $e^{-ik_3\tau_3}u_+(k_3 \tau_3)v_+(k_3 \tau_3)$ or $e^{+ik_3\tau_3}u_+^*(k_3 \tau_3)v_+^*(k_3 \tau_3)$. 
In both cases, the dominant terms in the expansion \eqref{eq: fermion Whittaker late} are the ones containing $(-k_3\tau_3)^{-i \mut}$, where the sign in the exponent agrees with the sign in the exponential $e^{-3ik_3 \tau_3}$.
The mode functions $u_-(k_3 \tau_3)$, $v_-(k_3\tau_3)$ do not contribute, because their prefactor of $(-k_3\tau_3)^{-i \mut}$ is exponentially suppressed for large $\lt\gg \mt$.
In conclusion, we need to perform the two following integrals over $x_3=k_3\tau_3$ (together with their conjugates):
\begin{gather}
\int_{-\infty}^0 \mathrm d x_3\, u_+(x_3) v_+(x_3) e^{-ix_3} =c_3 \,,\label{eq: int x3 0}\\
\int_{-\infty}^0 \mathrm d x_3\, u_+(x_3) v_+(x_3) e^{-ix_3} (ix_3) = c_3 \cdot \frac 13 (1-2i\mut) \,, \label{eq: int x3 1}\\
c_3 = \left[\frac{m e^{\pi(\lt-\mut)} \Gamma^2(2i\mut)}{\Gamma(1+i(\mut+\lt))\Gamma(i(\mut+\lt))} \right] \left(\frac 94\right)^{i\mut}\frac i3 e^{\pi\mut} \Gamma(1-2i\mut) \,.
\end{gather}
We notice that the factor in squared brackets in $c_3$ tends to $\exp\left(-2\pi(\mut-\lt)\right)\sim \exp\left( -\pi \mt^2/\lt\right)$ in the limit $\mt^2/\lt \to \infty$. 
This leads to the final exponential factor in $\fNL$ in the aforementioned limit, which was to be expected from the arguments exposed in Sec.~\ref{sec: fNL estimate}.
We also observe that the integral in Eq.~\eqref{eq: int x3 1} has an enhancement of $\mut$ with respect to the integral in \eqref{eq: int x3 0}, and gives the leading contribution. 

The remaining integrals in $\tau_1$, $\tau_2$ involve similar integrals as the ones collected in Eq.~\eqref{eq: int x3 0} and \eqref{eq: int x3 1}:
\begin{gather}
\int_{-\infty}^0 \mathrm d x_1\, (-2x_1)^{-i\mut} e^{-ix_1} =  2^{-i\mut}ie^{\pi\mut/2} \Gamma(1-i\mut)\,,\label{eq: int x1 0}\\
\int_{-\infty}^0 \mathrm d x_1\, (-2x_1)^{-i\mut} e^{-ix_1} (ix_1) =  2^{-i\mut}ie^{\pi\mut/2} \Gamma(1-i\mut)\cdot (1-i\mut) \,. \label{eq: int x1 1}
\end{gather}

We have now all the ingredients to perform the calculation of the amplitude. 
In the final expression, the dominant term in the limit $\lt\to\infty$ turns out to scale as $\lt^2 \mt^3$, in agreement with the expectations of Sec.~\eqref{sec: fNL estimate}, and contains an oscillating phase $(k_3/k_1)^{-2i\mut}$.

We can finally convert the 3-point function of the inflaton fluctuations $\delta\phi$ into the observable shape $S(k_1,k_2,k_3)$
\begin{gather}
\left \langle \zeta(\vk_1) \zeta(\vk_2) \zeta(\vk_3) \right \rangle =
  (2\pi)^3\delta^3 \left(\vec k_1+\vec k_2 +\vec k_3\right)
  \left \langle \zeta(\vk_1) \zeta(\vk_2) \zeta(\vk_3) \right \rangle^\prime \,, \label{eq: 3pt zeta prime}\\
\left \langle \zeta(\vk_1) \zeta(\vk_2) \zeta(\vk_3) \right \rangle^\prime= \frac{(2\pi)^4 \Pz^2}{k_1^2 k_2^2 k_3^2} S(k_1,k_2,k_3) \label{eq: def shape}
\end{gather}
of the 3-point function of the curvature perturbation $\zeta$, which in the flat gauge can be written as $\zeta =-H \delta\phi/\dot\phi$.

The final result that we obtain for the contribution to the squeezed shape from one SM fermion (accounting for two chiralities, and the respective color factor $N_c$) is, up to a constant phase,
\begin{equation}
\label{eq: shape}
\begin{aligned}
S(k_1,k_2,k_3) = & \frac{k_1^4 k_3^2}{(2\pi)^4 \Pz^2} 
  \left \langle
  \zeta(\vk_1) \zeta(\vk_2) \zeta(\vk_3)
  \right \rangle^\prime = 
  \frac{k_1^4 k_3^2}{(2\pi)^4 \Pz^2} 
  \left(\frac{H}{\dot\phi}\right)^3
  \left \langle
  \delta\phi(\vk_1) \delta\phi(\vk_2) \delta\phi(\vk_3)
  \right \rangle^\prime =\\
  \overset{\lambda \gg m }{\overset{k_3\ll k_1\sim k_2}{\simeq}} & \ 
  \frac{N_c}{6\pi} \Pz^{-1/2} 
  \left( \frac{m}{\Lf} \right)^3 
  \lt^2
  \frac{\mut e^{\pi \lt} \Gamma(-i\mut)^2 \Gamma(2i\mut)^3}{2\pi \Gamma\big(i(\mut+\lt)\big)^3 \Gamma\big(1+i(\mut-\lt) \big)}
  \left( \frac{k_3}{k_1}\right)^{2-2i\mu}
\end{aligned}
\end{equation}
The second to last term in Eq.~\eqref{eq: shape} tends to $\exp\left(-2\pi(\mut-\lt)\right)\sim \exp\left( -\pi \mt^2/\lt\right)$ in the limit $\lt\to\infty$, $\mt^2/\lt\to\infty$.

\section{Higgs potential in the early universe}\label{app:higgs}

In the early universe, the Higgs potential changes in several ways.  The main effect is that the Higgs mass can be changed drastically, an effect not un-related to the Electroweak Hierarchy problem. 
During inflation, higher dimensional operators, particle production, and a non-minimal coupling to gravity can all change the Higgs mass in an inflating background. 
During reheating, thermal corrections are crucial in bringing the Higgs field back to the electroweak minimum. In this Appendix, we discuss in more detail some of these effects.

\subsection{Higgs potential during inflation}\label{app:higgsinflation}

\paragraph{$\xi R \mathcal{H}^{\dagger}\mathcal{H}$ coupling} The Higgs Lagrangian in Eq.~\eqref{eq: Lagrangian Higgs} should include a non minimal coupling of the Higgs to the Ricci scalar $\xi R \mathcal{H}^{\dagger}\mathcal{H}$.
The order of magnitude of $\xi$, due to its RG flow, cannot be smaller than $10^{-2}$.
We choose a sign convention such that during inflation the contribution to the Higgs potential is $V_h\supset -6\xi H^2 h^2$.
By adding this term to the Lagrangian \eqref{eq: Lagrangian Higgs}, the vev of the Higgs field $h$ during inflation is
\begin{equation}
\vUV = \sqrt{\tfrac 43 \lUV}\LH \left(\frac 12 + \frac 12\sqrt{ 1 +\frac{36 \xi H^2}{\lUV^2\LH^2}} \right)^{1/2} =
\sqrt{\tfrac 43 \lUV}\LH  \left(1 + \frac{9 \xi H^2}{2\lUV^2\LH^2} + \mathcal O \left(\frac{\xi^2 H^4}{\LH^4}\right)\right) \,.
\label{eq: vev with xi}
\end{equation}
The impact of the non minimal coupling is small as long as 
\begin{equation}
\xi < \frac{\lUV^2\LH^2}{H^2} \overset{\text{Eq.~}\eqref{eq: constr thermal rescue}} < 10^2 \left(\frac{\lUV}{0.01}\right) \left(\frac{H}{6\cdot 10^{13}\GeV} \right)^{-1} \,.
\end{equation}
If $\xi< -\frac{1}{36}\frac{\lUV^2\LH^2}{H^2}$ the positive mass removes the UV minimum; using Eq.~\eqref{eq: constr thermal rescue}, this can be rephrased as saying that the UV minimum is removed if $\xi<-3\left(\tfrac{H}{6\cdot 10^{13}\GeV} \right)^{-1}\left(\tfrac{\lUV}{0.01} \right)$. 
Another effect of a large positive mass is that if $\xi<-3/16$ then the positive mass term is large enough to damp the quantum fluctuations and the Higgs field would not move from the origin during inflation. 

If $|\xi|\gtrsim\mathcal O(10)$, then the $\xi R$ term during preheating (when the inflaton rolls around the minimum) switches from positive to negative values at each oscillation, and can source a tachyonic instability for the Higgs \cite{Figueroa:2017slm}. 
Due to the assumption of instantaneous reheating, as well as the large initial field value $\vUV$, this issue is not important in our situation.

In summary, we can identify the following regimes of interest for $\xi$:
\begin{enumerate}
\item $\xi<-3/16$: the Higgs is stabilized at the origin, making it more likely that the Higgs is sitting at the origin instead of the UV minimum during inflation.
\item $-3/16 < \xi \lesssim 10^2\left(\tfrac{H}{6\cdot 10^{13}\GeV} \right)^{-1}\left(\tfrac{\lUV}{0.01} \right)$: the effect of the non minimal coupling is small.
\item $\xi \gtrsim 10^2\left(\tfrac{H}{6\cdot 10^{13}\GeV} \right)^{-1}\left(\tfrac{\lUV}{0.01} \right)$: $\vUV$ is brought to larger values, roughly at $\vUV^{(\xi)}\simeq 2\xi^{1/4} \sqrt{H \LH}$.
Given the milder dependence on $\LH$ than Eq.~\eqref{eq: vUV}, in this case the condition for the Higgs to be rescued by thermal effects becomes $\LH<\xi^{-1/2}\cdot 2 \cdot 10^{17}\GeV$.
\end{enumerate}
For definiteness, we assume that we are dealing with the second case, in which $\xi$ is irrelevant.
The third scenario of very large and positive $\xi$ can be more naturally described as a direct coupling between the Higgs and the inflaton.

\paragraph{Higgs inflaton coupling and particle production} The observability of the direct couplings $\mathcal O_{h1}$ and $\mathcal O_{h2}$ between the Higgs and the inflaton is studied in subsection~\ref{sec:otheroperator}. 
Here, we mainly discuss the effect of these operators on the Higgs potential. 

Such direct couplings between the inflaton and the Higgs can be generated by integrating out fermions who couple directly to the inflaton:
\begin{equation}
\mathcal{O}_{h1} \propto -\frac{\sum c_i^2 y_i^2 N_{c,\,i}}{16\pi^2 \Lambda_f^2} (\partial \phi)^2 \mathcal{H}^{\dagger} \mathcal{H},
\end{equation}
During inflation, this coupling would generate a correction to the Higgs potential proportional to $\lambda^2 \sim 10^2 H^2 \gg H^2$ for the parameter space of interest to us.
If the $c_i$ couplings are of comparable size, the top quark dominates and the correction to the Higgs mass would be $ \mu_h^2 \approx - \left(\frac{y_t^2 \lambda}{\pi}\right)^2$. 

Another important effect comes from particle production during inflation. 
The observable effects will be discussed in more detail in a companion paper \cite{Hook:2019}. 
The main effect on the Higgs potential comes from the same fermion condensate discussed in Sec.~\ref{sec: fNL estimate}. 
Unlike thermally produced particles, which generically lead to symmetry restoration, the fermions produced through the inflaton coupling will generate a negative squared mass term for the Higgs in the form of%
\footnote{We did not manage to compute the full integral in de Sitter space to reproduce the coefficient in front of $(y_i h)^2/\lambda_i H $ in the exponent, but since this exponential factor arises from the same particle production suppression as in the case of the 3-point function of the inflaton, we expect the final result to be similar.}
\begin{equation}
\mu_h^2 \approx - \frac{y_i^2 \lambda_i^2 N_{c,\, i}}{\pi^2} \exp\left[ - \frac{\pi (y_i h)^2}{\lambda_i H} \right].
\end{equation}
Such a mass term, in the limit where $\lambda \gg H$, will induce a Higgs vev during inflation that is of order $\sqrt{\lambda H} \gtrsim H^2$, which is one of the many reasons why we study predominantly the case where $\vUV \gg H$.

\subsection{Higgs dynamics during reheating}\label{app:higgsreheat}

In this subsection, we discuss the Higgs dynamics after the universe reaches the maximal temperature $T_\text{max}$. 
In absence of a concrete model for inflation, we stay agnostic about the mechanism that produces an instantaneous reheating, though it can easily be arranged by a waterfall field that couples strongly to the SM sector in hybrid inflation models~\cite{Linde:1993cn}. 
As explained in the main text, reheating generates a thermal bath of SM particles which contribute to the Higgs potential with a thermal mass  
\cite{Espinosa:2015qea,Ema:2016kpf,Kohri:2016wof,Enqvist:2016mqj,Postma:2017hbk,Ema:2017loe,Joti:2017fwe,Rajantie:2017ajw} 
\begin{equation}
V_T(h) \simeq \frac 12 \kappa T^2\,h^2 e^{-h/(2\pi T)}\,, \quad \kappa \simeq 0.12 \,.
\end{equation}
For $T_\text{max} \gtrsim \vUV$, the exponential factor is lifted and the UV minimum $\vUV$ of the Higgs potential disappears if
\begin{equation}
T_\text{max} \gtrsim \left|\frac{\lUV}{\kappa}\right|^{1/2} \vUV.
\end{equation}
For the small $\lUV\sim \mathcal O(10^{-2})$ at the energy scales relevant in this paper, this is automatically satisfied when $T_\text{max} \gtrsim \vUV$.
This is the condition shown in Fig.~\ref{fig: plane v H} as a constraint on the parameter space (infering the relation between $H$ and $T_\text{max}$ by assuming instantaneous reheating).

After the temperature of the SM bath has reached $T_\text{max}$, the Higgs field starts to oscillate around $h=0$ while the amplitude of the oscillations decreases due to the Hubble friction and interactions with the SM thermal bath~\cite{Elmfors:1993re}. 
The Higgs field redshifts like radiation both when the potential is dominated by the thermal correction, and when the field value is small enough that the quartic coupling is positive and dominates the Higgs potential. 
In both cases, the ratio between the amplitude of the Higgs oscillation and the temperature of the thermal bath remains constant, which ensures that the Higgs background finally lands in the electroweak minimum. 

The Higgs oscillation amplitude also decreases as a result of its interaction with the SM bath or simply the decay into lighter particles. 
This decay of the amplitude of the Higgs field is dominated by its interaction with the electroweak gauge bosons in the thermal bath. 
The rate $\Gamma_h$ can be found to be~\cite{Elmfors:1993re}
\begin{equation}
\Gamma_h = \frac{3 g_2^4 T^2}{256 \pi m_T} \approx 10^{-3} T
\end{equation}
where $m_T^2 = \kappa T^2$ is the thermal mass of the Higgs. This rate becomes faster than the Hubble rate soon after reheating, so that the amplitude of the oscillations of the Higgs background decays very quickly and the Higgs sits at the electroweak minimum soon after reheating.

\section{Inflaton couplings and two point function}\label{app:inflaton}

In this Appendix, we discuss the relation between the couplings $c_i$ of the inflaton to the SM fermions from a UV perspective, assessing the implications for the 2-point function of the inflaton. 
We also summarize some other consistency conditions that the parameters need to satisfy.

In principle, the inflaton can couple to each individual SM fermion independently, in the form of
\begin{equation}
\mathcal{L} \supset \partial_\mu \phi J_{\phi}^{\mu} =  \partial_\mu \phi \sum_{F_i} \frac{c_{F_i}}{\Lf} F_i^{\dagger} \overline \sigma^{\mu} F_i \,, 
\end{equation}
where $F_i = Q,\, u^c,\, d^c,\, L,\, e^c$ are the SM fermions in two component notation with charges under the SM gauge group collected in Table~\ref{tab:U1 charges}. 
However, if the current $J_\phi^\mu$ corresponds to a $U(1)$ symmetry that is anomalous, either the SM gauge group is broken, or the inflaton $\phi$ would receive a UV sensitive correction to its kinetic term from the 3-loop diagram
\begin{equation}
\frac{c_{F_i}^2}{16\pi^2} \left(\frac{g_j^2}{16\pi^2}\right)\left(\frac{g_k^2}{16\pi^2}\right) \left(\frac{\Lambda_{*}}{\Lf}\right)^2 \left(\partial \phi\right)^2,
\end{equation}
where $g_j$ and $g_k$ are SM gauge couplings. 
This is analogous to the minimal mass gauge bosons receive if the symmetry is anomalous~\cite{Preskill:1990fr}. 
Such a contribution can change the dynamics of the inflaton if the ratio between the cutoff $\Lambda_*$ and the scale $\Lf$ is significant%
\footnote{In reality, this might simply mean that there cannot be too much axion monodromy~\cite{McAllister:2008hb} or clock-working~\cite{Kaplan:2015fuy}, which is not necessarily a requirement.}. 
The absence of an anomaly also ensures that there is no significant production of gauge bosons through the coupling $ \frac{\phi}{\Lambda_G} G\tilde{G}$.

The only universal (i.~e.~flavour independent) anomaly-free $U(1)$ extensions of the Standard Model are $U(1)_Y$ and $U(1)_{B-L}$ and their linear combinations $U(1)'$. The charges of the SM fermions under $U(1)'$ are given in table~\ref{tab:U1 charges}, while the corresponding coefficients of the vector and axial vector currents for the fermions are reported in Table~\ref{tab:U1 coefficients}. 
The choice we made in the main text of the paper where all $c_i$'s are identical up to a sign for each individual family, is, as a result, a point that is preferred by UV considerations.

\begin{table}[h!]
\centering
\begin{tabular}{cccccc}
\toprule
 & $SU(3)$ & $SU(2)$ & $U(1)_Y$  & $U(1)_{B-L}$  & $U(1)'$ \\
\midrule
$Q = \begin{pmatrix} u\\ d \end{pmatrix}$ & $\textbf{3}$ & $\textbf{2}$ & $\tfrac 16$ & $+\tfrac 13$ & $\frac 16 \cth + \frac 13 \sth$\smallskip\\
$u^c$ & $\overline{\textbf{3}}$ & $\textbf{1}$ & $-\tfrac 23$ & $-\tfrac 13$ & $-\tfrac 23 \cth -\frac 13\sth$\smallskip\\
$d^c$ & $\overline{\textbf{3}}$ & $\textbf{1}$ & $\tfrac 13$ & $-\tfrac 13$ & $\tfrac 13 \cth - \frac 13 \sth$\smallskip \\
$L = \begin{pmatrix} \nu \\ e \end{pmatrix}$ & $\textbf{1}$ & $\textbf{2}$ & $-\tfrac 12$ & $-1$ & $-\frac 12 \cth -\sth$ \smallskip\\
$e^c$ & $\textbf{1}$ & $\textbf{1}$ & $1$ & $+1$ & $\cth+\sth$ \smallskip\\
\bottomrule
\end{tabular}
\caption{Charges of the SM fermions content under $U(1)' \equiv \left( \cth \, U(1)_Y +\sth \, U(1)_{B-L} \right)$ in two component spinor notation.}
\label{tab:U1 charges}
\end{table}

\begin{table}[h!]
\centering
\begin{tabular}{ccc}
\toprule
SM fermion $f$ & Vector current & Axial current \\
\midrule
up quarks & $\frac{5}{12} \cth +\frac 13 \sth$ & $\frac 14 \cth$ \\
down quarks & $-\frac{1}{12} \cth +\frac 13 \sth$ & $-\frac 14 \cth$ \\ 
leptons & $- \frac 34 \cth -\sth$ & $-\frac 14 \cth$ \\
\bottomrule
\end{tabular}
\caption{Coefficients of the vector and axial vector bilinear currents for the SM fermions $f$ (in four component notation). The coefficients $c_V$ (vector) and $c_A$ (axial) are obtained from the coefficients of the left handed ($c_L$) and right handed ($c_R$) fermions in the SM via $c_L P_L +c_R P_R = \tfrac{c_L+c_R}{2} + \tfrac{-c_L+c_R}{2}\gamma_5 \Rightarrow c_V= \tfrac{c_L+c_R}{2},\, c_A=\tfrac{-c_L+c_R}{2}$.}
\label{tab:U1 coefficients}
\end{table}

In the end, let us comment on the effect of the fermion density on the motion of the inflaton. The fermion density induces a correction to the equation of motion for the inflaton \cite{Adshead:2018oaa}:
\begin{equation}
\ddot{\phi} + 3 H \dot{\phi} = - V'(\phi) + \frac{c_{f_i} m_{f_i}}{\Lf} \overline{f} \gamma^5 f
\end{equation}
In order for the fermions to not significantly affect the dynamics of the inflaton, the following requirement needs to be satisfied
\begin{equation}
 H \dot{\phi} \gtrsim \frac{c_{f_i} m_{f_i}}{\Lambda_f} \overline{f} \gamma^5 f \approx \frac{c_{f_i} m_{f_i}^2 \lambda_i^2}{\Lambda_f} \exp[-\pi m_{f_i}^2/\lambda_i H],
\end{equation}
which is equivalent to the requirement
\begin{equation}
c_{f_i}^2 \left(\frac{c_{f_i} \dot{\phi}}{\Lf^2}\right)^2 \left(\frac{m_{f_i}^2}{\lambda_i H} \exp[-\pi m_{f_i}^2/\lambda_i H]\right)\lesssim 1
\end{equation}
This is easily satisfied as both terms in brackets are smaller than 1.

One additional worry regarding the fermion density is the annihilation of the fermions into massless gauge bosons or lighter fermions right when they are produced. 
Annihilations into massless gauge bosons or fermions with a normal dispersion relation can only happen between fermions whose spatial momenta are nearly opposite with modulus $|{\vec k}| \approx \lambda \gg m$. 
Therefore, the annihilation rate is $\mathcal{O}(g_i^4 m/4\pi)$, much smaller than the Hubble expansion rate. 

Similarly, annihilations into lighter SM fermions with a dispersion similar to the one of the annihilating fermions has a cross section
\begin{equation}
\sigma \sim \frac{g_i^4}{4\pi ({\vec k_1}-{\vec k_2})^2} \approx \frac{g_i^4}{4\pi {\lambda}^2}
\end{equation}
if the fermions were not exactly back to back. 
This suggest an annihilation rate into light fermions similar to the one into gauge bosons, and therefore should also be negligible. 
In reality, since the lighter fermions that the heavy fermions can annihilate to are likely also produced with a high density, Fermi-blocking can forbid this annihilation entirely, for the same reason why the $t$-channel scattering of fermions cannot thermalize a Fermi-degenerate gas.
\bibliographystyle{JHEP}
\bibliography{newhvev_Refs}

\providecommand{\href}[2]{#2}\begingroup\raggedright\begin{thebibliography}{10}

\bibitem{Sher:1988mj}
M.~Sher, {\it {Electroweak Higgs Potentials and Vacuum Stability}},  {\em Phys.
  Rept.} {\bf 179} (1989) 273--418.

\bibitem{Arnold:1989cb}
P.~B. Arnold, {\it {Can the Electroweak Vacuum Be Unstable?}},  {\em Phys.
  Rev.} {\bf D40} (1989) 613.

\bibitem{Altarelli:1994rb}
G.~Altarelli and G.~Isidori, {\it {Lower limit on the Higgs mass in the
  standard model: An Update}},  {\em Phys. Lett.} {\bf B337} (1994) 141--144.

\bibitem{Casas:1996aq}
J.~A. Casas, J.~R. Espinosa, and M.~Quiros, {\it {Standard model stability
  bounds for new physics within LHC reach}},  {\em Phys. Lett.} {\bf B382}
  (1996) 374--382, [\href{http://arxiv.org/abs/hep-ph/9603227}{{\tt
  hep-ph/9603227}}].

\bibitem{Hambye:1996wb}
T.~Hambye and K.~Riesselmann, {\it {Matching conditions and Higgs mass upper
  bounds revisited}},  {\em Phys. Rev.} {\bf D55} (1997) 7255--7262,
  [\href{http://arxiv.org/abs/hep-ph/9610272}{{\tt hep-ph/9610272}}].

\bibitem{Isidori:2001bm}
G.~Isidori, G.~Ridolfi, and A.~Strumia, {\it {On the metastability of the
  standard model vacuum}},  {\em Nucl. Phys.} {\bf B609} (2001) 387--409,
  [\href{http://arxiv.org/abs/hep-ph/0104016}{{\tt hep-ph/0104016}}].

\bibitem{Ellis:2009tp}
J.~Ellis, J.~R. Espinosa, G.~F. Giudice, A.~Hoecker, and A.~Riotto, {\it {The
  Probable Fate of the Standard Model}},  {\em Phys. Lett.} {\bf B679} (2009)
  369--375, [\href{http://arxiv.org/abs/0906.0954}{{\tt arXiv:0906.0954}}].

\bibitem{Bezrukov:2012sa}
F.~Bezrukov, M.~{\relax Yu}. Kalmykov, B.~A. Kniehl, and M.~Shaposhnikov, {\it
  {Higgs Boson Mass and New Physics}},  {\em JHEP} {\bf 10} (2012) 140,
  [\href{http://arxiv.org/abs/1205.2893}{{\tt arXiv:1205.2893}}]. [,275(2012)].

\bibitem{Bednyakov:2015sca}
A.~V. Bednyakov, B.~A. Kniehl, A.~F. Pikelner, and O.~L. Veretin, {\it
  {Stability of the Electroweak Vacuum: Gauge Independence and Advanced
  Precision}},  {\em Phys. Rev. Lett.} {\bf 115} (2015), no.~20 201802,
  [\href{http://arxiv.org/abs/1507.08833}{{\tt arXiv:1507.08833}}].

\bibitem{EliasMiro:2011aa}
J.~Elias-Miro, J.~R. Espinosa, G.~F. Giudice, G.~Isidori, A.~Riotto, and
  A.~Strumia, {\it {Higgs mass implications on the stability of the electroweak
  vacuum}},  {\em Phys. Lett.} {\bf B709} (2012) 222--228,
  [\href{http://arxiv.org/abs/1112.3022}{{\tt arXiv:1112.3022}}].

\bibitem{Degrassi:2012ry}
G.~Degrassi, S.~Di~Vita, J.~Elias-Miro, J.~R. Espinosa, G.~F. Giudice,
  G.~Isidori, and A.~Strumia, {\it {Higgs mass and vacuum stability in the
  Standard Model at NNLO}},  {\em JHEP} {\bf 08} (2012) 098,
  [\href{http://arxiv.org/abs/1205.6497}{{\tt arXiv:1205.6497}}].

\bibitem{Buttazzo:2013uya}
D.~Buttazzo, G.~Degrassi, P.~P. Giardino, G.~F. Giudice, F.~Sala, A.~Salvio,
  and A.~Strumia, {\it {Investigating the near-criticality of the Higgs
  boson}},  {\em JHEP} {\bf 12} (2013) 089,
  [\href{http://arxiv.org/abs/1307.3536}{{\tt arXiv:1307.3536}}].

\bibitem{Guth:1980zm}
A.~H. Guth, {\it {The Inflationary Universe: A Possible Solution to the Horizon
  and Flatness Problems}},  {\em Phys. Rev.} {\bf D23} (1981) 347--356. [Adv.
  Ser. Astrophys. Cosmol.3,139(1987)].

\bibitem{Starobinsky:1980te}
A.~A. Starobinsky, {\it {A New Type of Isotropic Cosmological Models Without
  Singularity}},  {\em Phys. Lett.} {\bf B91} (1980) 99--102. [,771(1980)].

\bibitem{Linde:1981mu}
A.~D. Linde, {\it {A New Inflationary Universe Scenario: A Possible Solution of
  the Horizon, Flatness, Homogeneity, Isotropy and Primordial Monopole
  Problems}},  {\em Phys. Lett.} {\bf 108B} (1982) 389--393. [Adv. Ser.
  Astrophys. Cosmol.3,149(1987)].

\bibitem{Mukhanov:1981xt}
V.~F. Mukhanov and G.~V. Chibisov, {\it {Quantum Fluctuations and a Nonsingular
  Universe}},  {\em JETP Lett.} {\bf 33} (1981) 532--535. [Pisma Zh. Eksp.
  Teor. Fiz.33,549(1981)].

\bibitem{Albrecht:1982wi}
A.~Albrecht and P.~J. Steinhardt, {\it {Cosmology for Grand Unified Theories
  with Radiatively Induced Symmetry Breaking}},  {\em Phys. Rev. Lett.} {\bf
  48} (1982) 1220--1223. [Adv. Ser. Astrophys. Cosmol.3,158(1987)].

\bibitem{Lyth:1998xn}
D.~H. Lyth and A.~Riotto, {\it {Particle physics models of inflation and the
  cosmological density perturbation}},  {\em Phys. Rept.} {\bf 314} (1999)
  1--146, [\href{http://arxiv.org/abs/hep-ph/9807278}{{\tt hep-ph/9807278}}].

\bibitem{Baumann:2009ds}
D.~Baumann, {\it {Inflation}},  in {\em {Physics of the large and the small,
  TASI 09, proceedings of the Theoretical Advanced Study Institute in
  Elementary Particle Physics, Boulder, Colorado, USA, 1-26 June 2009}},
  pp.~523--686, 2011.
\newblock \href{http://arxiv.org/abs/0907.5424}{{\tt arXiv:0907.5424}}.

\bibitem{Espinosa:2007qp}
J.~R. Espinosa, G.~F. Giudice, and A.~Riotto, {\it {Cosmological implications
  of the Higgs mass measurement}},  {\em JCAP} {\bf 0805} (2008) 002,
  [\href{http://arxiv.org/abs/0710.2484}{{\tt arXiv:0710.2484}}].

\bibitem{Kobakhidze:2013tn}
A.~Kobakhidze and A.~Spencer-Smith, {\it {Electroweak Vacuum (In)Stability in
  an Inflationary Universe}},  {\em Phys. Lett.} {\bf B722} (2013) 130--134,
  [\href{http://arxiv.org/abs/1301.2846}{{\tt arXiv:1301.2846}}].

\bibitem{Kobakhidze:2014xda}
A.~Kobakhidze and A.~Spencer-Smith, {\it {The Higgs vacuum is unstable}},
  \href{http://arxiv.org/abs/1404.4709}{{\tt arXiv:1404.4709}}.

\bibitem{Fairbairn:2014zia}
M.~Fairbairn and R.~Hogan, {\it {Electroweak Vacuum Stability in light of
  BICEP2}},  {\em Phys. Rev. Lett.} {\bf 112} (2014) 201801,
  [\href{http://arxiv.org/abs/1403.6786}{{\tt arXiv:1403.6786}}].

\bibitem{Enqvist:2014bua}
K.~Enqvist, T.~Meriniemi, and S.~Nurmi, {\it {Higgs Dynamics during
  Inflation}},  {\em JCAP} {\bf 1407} (2014) 025,
  [\href{http://arxiv.org/abs/1404.3699}{{\tt arXiv:1404.3699}}].

\bibitem{Hook:2014uia}
A.~Hook, J.~Kearney, B.~Shakya, and K.~M. Zurek, {\it {Probable or Improbable
  Universe? Correlating Electroweak Vacuum Instability with the Scale of
  Inflation}},  {\em JHEP} {\bf 01} (2015) 061,
  [\href{http://arxiv.org/abs/1404.5953}{{\tt arXiv:1404.5953}}].

\bibitem{Kamada:2014ufa}
K.~Kamada, {\it {Inflationary cosmology and the standard model Higgs with a
  small Hubble induced mass}},  {\em Phys. Lett.} {\bf B742} (2015) 126--135,
  [\href{http://arxiv.org/abs/1409.5078}{{\tt arXiv:1409.5078}}].

\bibitem{Shkerin:2015exa}
A.~Shkerin and S.~Sibiryakov, {\it {On stability of electroweak vacuum during
  inflation}},  {\em Phys. Lett.} {\bf B746} (2015) 257--260,
  [\href{http://arxiv.org/abs/1503.02586}{{\tt arXiv:1503.02586}}].

\bibitem{Kearney:2015vba}
J.~Kearney, H.~Yoo, and K.~M. Zurek, {\it {Is a Higgs Vacuum Instability Fatal
  for High-Scale Inflation?}},  {\em Phys. Rev.} {\bf D91} (2015), no.~12
  123537, [\href{http://arxiv.org/abs/1503.05193}{{\tt arXiv:1503.05193}}].

\bibitem{East:2016anr}
W.~E. East, J.~Kearney, B.~Shakya, H.~Yoo, and K.~M. Zurek, {\it {Spacetime
  Dynamics of a Higgs Vacuum Instability During Inflation}},  {\em Phys. Rev.}
  {\bf D95} (2017), no.~2 023526, [\href{http://arxiv.org/abs/1607.00381}{{\tt
  arXiv:1607.00381}}].

\bibitem{Herranen:2014cua}
M.~Herranen, T.~Markkanen, S.~Nurmi, and A.~Rajantie, {\it {Spacetime curvature
  and the Higgs stability during inflation}},  {\em Phys. Rev. Lett.} {\bf 113}
  (2014), no.~21 211102, [\href{http://arxiv.org/abs/1407.3141}{{\tt
  arXiv:1407.3141}}].

\bibitem{Abazajian:2016yjj}
{\bf CMB-S4} Collaboration, K.~N. Abazajian et~al., {\it {CMB-S4 Science Book,
  First Edition}},  \href{http://arxiv.org/abs/1610.02743}{{\tt
  arXiv:1610.02743}}.

\bibitem{Chen:2009zp}
X.~Chen and Y.~Wang, {\it {Quasi-Single Field Inflation and
  Non-Gaussianities}},  {\em JCAP} {\bf 1004} (2010) 027,
  [\href{http://arxiv.org/abs/0911.3380}{{\tt arXiv:0911.3380}}].

\bibitem{Baumann:2011nk}
D.~Baumann and D.~Green, {\it {Signatures of Supersymmetry from the Early
  Universe}},  {\em Phys. Rev.} {\bf D85} (2012) 103520,
  [\href{http://arxiv.org/abs/1109.0292}{{\tt arXiv:1109.0292}}].

\bibitem{Arkani-Hamed:2015bza}
N.~Arkani-Hamed and J.~Maldacena, {\it {Cosmological Collider Physics}},
  \href{http://arxiv.org/abs/1503.08043}{{\tt arXiv:1503.08043}}.

\bibitem{Lee:2016vti}
H.~Lee, D.~Baumann, and G.~L. Pimentel, {\it {Non-Gaussianity as a Particle
  Detector}},  {\em JHEP} {\bf 12} (2016) 040,
  [\href{http://arxiv.org/abs/1607.03735}{{\tt arXiv:1607.03735}}].

\bibitem{Chen:2016hrz}
X.~Chen, Y.~Wang, and Z.-Z. Xianyu, {\it {Standard Model Mass Spectrum in
  Inflationary Universe}},  {\em JHEP} {\bf 04} (2017) 058,
  [\href{http://arxiv.org/abs/1612.08122}{{\tt arXiv:1612.08122}}].

\bibitem{Meerburg:2016zdz}
P.~D. Meerburg, M.~Münchmeyer, J.~B. Muñoz, and X.~Chen, {\it {Prospects for
  Cosmological Collider Physics}},  {\em JCAP} {\bf 1703} (2017), no.~03 050,
  [\href{http://arxiv.org/abs/1610.06559}{{\tt arXiv:1610.06559}}].

\bibitem{Kumar:2017ecc}
S.~Kumar and R.~Sundrum, {\it {Heavy-Lifting of Gauge Theories By Cosmic
  Inflation}},  {\em JHEP} {\bf 05} (2018) 011,
  [\href{http://arxiv.org/abs/1711.03988}{{\tt arXiv:1711.03988}}].

\bibitem{Anber:2009ua}
M.~M. Anber and L.~Sorbo, {\it {Naturally inflating on steep potentials through
  electromagnetic dissipation}},  {\em Phys. Rev.} {\bf D81} (2010) 043534,
  [\href{http://arxiv.org/abs/0908.4089}{{\tt arXiv:0908.4089}}].

\bibitem{Chen:2018xck}
X.~Chen, Y.~Wang, and Z.-Z. Xianyu, {\it {Neutrino Signatures in Primordial
  Non-Gaussianities}},  {\em JHEP} {\bf 09} (2018) 022,
  [\href{http://arxiv.org/abs/1805.02656}{{\tt arXiv:1805.02656}}].

\bibitem{Adshead:2018oaa}
P.~Adshead, L.~Pearce, M.~Peloso, M.~A. Roberts, and L.~Sorbo, {\it
  {Phenomenology of fermion production during axion inflation}},  {\em JCAP}
  {\bf 1806} (2018), no.~06 020, [\href{http://arxiv.org/abs/1803.04501}{{\tt
  arXiv:1803.04501}}].

\bibitem{Weinberg:2008hq}
S.~Weinberg, {\it {Effective Field Theory for Inflation}},  {\em Phys. Rev.}
  {\bf D77} (2008) 123541, [\href{http://arxiv.org/abs/0804.4291}{{\tt
  arXiv:0804.4291}}].

\bibitem{Cheung:2007st}
C.~Cheung, P.~Creminelli, A.~L. Fitzpatrick, J.~Kaplan, and L.~Senatore, {\it
  {The Effective Field Theory of Inflation}},  {\em JHEP} {\bf 03} (2008) 014,
  [\href{http://arxiv.org/abs/0709.0293}{{\tt arXiv:0709.0293}}].

\bibitem{Akrami:2018odb}
{\bf Planck} Collaboration, Y.~Akrami et~al., {\it {Planck 2018 results. X.
  Constraints on inflation}},  \href{http://arxiv.org/abs/1807.06211}{{\tt
  arXiv:1807.06211}}.

\bibitem{Franciolini:2018ebs}
G.~Franciolini, G.~F. Giudice, D.~Racco, and A.~Riotto, {\it {Implications of
  the detection of primordial gravitational waves for the Standard Model}},
  \href{http://arxiv.org/abs/1811.08118}{{\tt arXiv:1811.08118}}.

\bibitem{Hook:2019}
A.~Hook, J.~Huang, and D.~Racco, {\it {Minimal signatures of the Standard Model
  in non-Gaussianities, to appear}},
  \href{http://arxiv.org/abs/1908.XXXX}{{\tt arXiv:1908.XXXX}}.

\bibitem{Espinosa:2015qea}
J.~R. Espinosa, G.~F. Giudice, E.~Morgante, A.~Riotto, L.~Senatore, A.~Strumia,
  and N.~Tetradis, {\it {The cosmological Higgstory of the vacuum
  instability}},  {\em JHEP} {\bf 09} (2015) 174,
  [\href{http://arxiv.org/abs/1505.04825}{{\tt arXiv:1505.04825}}].

\bibitem{Ema:2016kpf}
Y.~Ema, K.~Mukaida, and K.~Nakayama, {\it {Fate of Electroweak Vacuum during
  Preheating}},  {\em JCAP} {\bf 1610} (2016), no.~10 043,
  [\href{http://arxiv.org/abs/1602.00483}{{\tt arXiv:1602.00483}}].

\bibitem{Kohri:2016wof}
K.~Kohri and H.~Matsui, {\it {Higgs vacuum metastability in primordial
  inflation, preheating, and reheating}},  {\em Phys. Rev.} {\bf D94} (2016),
  no.~10 103509, [\href{http://arxiv.org/abs/1602.02100}{{\tt
  arXiv:1602.02100}}].

\bibitem{Enqvist:2016mqj}
K.~Enqvist, M.~Karciauskas, O.~Lebedev, S.~Rusak, and M.~Zatta, {\it
  {Postinflationary vacuum instability and Higgs-inflaton couplings}},  {\em
  JCAP} {\bf 1611} (2016) 025, [\href{http://arxiv.org/abs/1608.08848}{{\tt
  arXiv:1608.08848}}].

\bibitem{Postma:2017hbk}
M.~Postma and J.~van~de Vis, {\it {Electroweak stability and non-minimal
  coupling}},  {\em JCAP} {\bf 1705} (2017), no.~05 004,
  [\href{http://arxiv.org/abs/1702.07636}{{\tt arXiv:1702.07636}}].

\bibitem{Ema:2017loe}
Y.~Ema, M.~Karciauskas, O.~Lebedev, and M.~Zatta, {\it {Early Universe Higgs
  dynamics in the presence of the Higgs-inflaton and non-minimal Higgs-gravity
  couplings}},  {\em JCAP} {\bf 1706} (2017), no.~06 054,
  [\href{http://arxiv.org/abs/1703.04681}{{\tt arXiv:1703.04681}}].

\bibitem{Joti:2017fwe}
A.~Joti, A.~Katsis, D.~Loupas, A.~Salvio, A.~Strumia, N.~Tetradis, and
  A.~Urbano, {\it {(Higgs) vacuum decay during inflation}},  {\em JHEP} {\bf
  07} (2017) 058, [\href{http://arxiv.org/abs/1706.00792}{{\tt
  arXiv:1706.00792}}].

\bibitem{Rajantie:2017ajw}
A.~Rajantie and S.~Stopyra, {\it {Standard Model vacuum decay in a de Sitter
  Background}},  {\em Phys. Rev.} {\bf D97} (2018), no.~2 025012,
  [\href{http://arxiv.org/abs/1707.09175}{{\tt arXiv:1707.09175}}].

\bibitem{Barnaby:2010vf}
N.~Barnaby and M.~Peloso, {\it {Large Nongaussianity in Axion Inflation}},
  {\em Phys. Rev. Lett.} {\bf 106} (2011) 181301,
  [\href{http://arxiv.org/abs/1011.1500}{{\tt arXiv:1011.1500}}].

\bibitem{Barnaby:2011qe}
N.~Barnaby, E.~Pajer, and M.~Peloso, {\it {Gauge Field Production in Axion
  Inflation: Consequences for Monodromy, non-Gaussianity in the CMB, and
  Gravitational Waves at Interferometers}},  {\em Phys. Rev.} {\bf D85} (2012)
  023525, [\href{http://arxiv.org/abs/1110.3327}{{\tt arXiv:1110.3327}}].

\bibitem{ArkaniHamed:2004yi}
N.~Arkani-Hamed, S.~Dimopoulos, G.~F. Giudice, and A.~Romanino, {\it {Aspects
  of split supersymmetry}},  {\em Nucl. Phys.} {\bf B709} (2005) 3--46,
  [\href{http://arxiv.org/abs/hep-ph/0409232}{{\tt hep-ph/0409232}}].

\bibitem{Arvanitaki:2012ps}
A.~Arvanitaki, N.~Craig, S.~Dimopoulos, and G.~Villadoro, {\it {Mini-Split}},
  {\em JHEP} {\bf 02} (2013) 126, [\href{http://arxiv.org/abs/1210.0555}{{\tt
  arXiv:1210.0555}}].

\bibitem{Georgi:1974sy}
H.~Georgi and S.~L. Glashow, {\it {Unity of All Elementary Particle Forces}},
  {\em Phys. Rev. Lett.} {\bf 32} (1974) 438--441.

\bibitem{Pati:1974yy}
J.~C. Pati and A.~Salam, {\it {Lepton Number as the Fourth Color}},  {\em Phys.
  Rev.} {\bf D10} (1974) 275--289. [Erratum: Phys. Rev.D11,703(1975)].

\bibitem{Dimopoulos:1981zb}
S.~Dimopoulos and H.~Georgi, {\it {Softly Broken Supersymmetry and SU(5)}},
  {\em Nucl. Phys.} {\bf B193} (1981) 150--162.

\bibitem{Dimopoulos:1981yj}
S.~Dimopoulos, S.~Raby, and F.~Wilczek, {\it {Supersymmetry and the Scale of
  Unification}},  {\em Phys. Rev.} {\bf D24} (1981) 1681--1683.

\bibitem{Munoz:2015eqa}
J.~B. Muñoz, Y.~Ali-Haïmoud, and M.~Kamionkowski, {\it {Primordial
  non-gaussianity from the bispectrum of 21-cm fluctuations in the dark ages}},
   {\em Phys. Rev.} {\bf D92} (2015), no.~8 083508,
  [\href{http://arxiv.org/abs/1506.04152}{{\tt arXiv:1506.04152}}].

\bibitem{Alvarez:2014vva}
M.~Alvarez et~al., {\it {Testing Inflation with Large Scale Structure:
  Connecting Hopes with Reality}},  \href{http://arxiv.org/abs/1412.4671}{{\tt
  arXiv:1412.4671}}.

\bibitem{Shandera:2010ei}
S.~Shandera, N.~Dalal, and D.~Huterer, {\it {A generalized local ansatz and its
  effect on halo bias}},  {\em JCAP} {\bf 1103} (2011) 017,
  [\href{http://arxiv.org/abs/1010.3722}{{\tt arXiv:1010.3722}}].

\bibitem{Weinberg:2005vy}
S.~Weinberg, {\it {Quantum contributions to cosmological correlations}},  {\em
  Phys. Rev.} {\bf D72} (2005) 043514,
  [\href{http://arxiv.org/abs/hep-th/0506236}{{\tt hep-th/0506236}}].

\bibitem{Chen:2010xka}
X.~Chen, {\it {Primordial Non-Gaussianities from Inflation Models}},  {\em Adv.
  Astron.} {\bf 2010} (2010) 638979,
  [\href{http://arxiv.org/abs/1002.1416}{{\tt arXiv:1002.1416}}].

\bibitem{Chen:2017ryl}
X.~Chen, Y.~Wang, and Z.-Z. Xianyu, {\it {Schwinger-Keldysh Diagrammatics for
  Primordial Perturbations}},  {\em JCAP} {\bf 1712} (2017), no.~12 006,
  [\href{http://arxiv.org/abs/1703.10166}{{\tt arXiv:1703.10166}}].

\bibitem{dlmf}
NIST, ``{Digital Library of Mathematical functions}.''
  \url{https://dlmf.nist.gov/13#PT3}, 2010.

\bibitem{Figueroa:2017slm}
D.~G. Figueroa, A.~Rajantie, and F.~Torrenti, {\it {Higgs field-curvature
  coupling and postinflationary vacuum instability}},  {\em Phys. Rev.} {\bf
  D98} (2018), no.~2 023532, [\href{http://arxiv.org/abs/1709.00398}{{\tt
  arXiv:1709.00398}}].

\bibitem{Linde:1993cn}
A.~D. Linde, {\it {Hybrid inflation}},  {\em Phys. Rev.} {\bf D49} (1994)
  748--754, [\href{http://arxiv.org/abs/astro-ph/9307002}{{\tt
  astro-ph/9307002}}].

\bibitem{Elmfors:1993re}
P.~Elmfors, K.~Enqvist, and I.~Vilja, {\it {Thermalization of the Higgs field
  at the electroweak phase transition}},  {\em Nucl. Phys.} {\bf B412} (1994)
  459--478, [\href{http://arxiv.org/abs/hep-ph/9307210}{{\tt hep-ph/9307210}}].

\bibitem{Preskill:1990fr}
J.~Preskill, {\it {Gauge anomalies in an effective field theory}},  {\em Annals
  Phys.} {\bf 210} (1991) 323--379.

\bibitem{McAllister:2008hb}
L.~McAllister, E.~Silverstein, and A.~Westphal, {\it {Gravity Waves and Linear
  Inflation from Axion Monodromy}},  {\em Phys. Rev.} {\bf D82} (2010) 046003,
  [\href{http://arxiv.org/abs/0808.0706}{{\tt arXiv:0808.0706}}].

\bibitem{Kaplan:2015fuy}
D.~E. Kaplan and R.~Rattazzi, {\it {Large field excursions and approximate
  discrete symmetries from a clockwork axion}},  {\em Phys. Rev.} {\bf D93}
  (2016), no.~8 085007, [\href{http://arxiv.org/abs/1511.01827}{{\tt
  arXiv:1511.01827}}].

\end{thebibliography}\endgroup

\end{document}